\documentclass[bib]{statapress}

\usepackage[crop,newcenter,frame]{pagedims}

\usepackage{sj}

\usepackage{stata}

\usepackage{shadow}

\sjsetissue{$vv$}{$ii$}{$mm$}{$yyyy$}

\usepackage{amssymb,amsmath,bbm,mathrsfs,longtable,graphicx,subcaption}
\usepackage{hyperref}
\hypersetup{colorlinks=true,       
    linkcolor=blue,          
    citecolor=blue,        
    filecolor=magenta,      
    urlcolor=blue           
}
\renewcommand{\P}{\mathbbm{P}}
\newcommand{\E}{\mathbb{E}}
\newcommand{\V}{\mathbb{V}}
\newcommand{\I}{\mathbbm{1}}

\DeclareMathOperator*{\argmin}{arg\,min}

\newcommand{\cval}{\mathfrak{c}}

\newcommand{\bb}{\mathbf{b}}
\newcommand{\bD}{\mathbf{D}}
\newcommand{\bN}{\mathbf{N}}
\newcommand{\bQ}{\mathbf{Q}}
\newcommand{\bw}{\mathbf{w}}

\newcommand{\bbeta}{\boldsymbol{\beta}}
\newcommand{\btheta}{\boldsymbol{\theta}}
\newcommand{\bgamma}{\boldsymbol{\gamma}}

\newcommand{\bSigma}{\boldsymbol{\Sigma}}

\begin{document}

\inserttype[st0001]{article}

\title[Binscatter]{Binscatter Regressions}

\author{M. D. Cattaneo, R. K. Crump, M. H. Farrell, and Y. Feng}{
  Matias D. Cattaneo\\Princeton University\\Princeton, NJ\\cattaneo@princeton.edu
  \and
  Richard K. Crump\\Federal Reserve Bank of New York\\New York, NY\\richard.crump@ny.frb.org\\ $ $
  \and
  Max H. Farrell\\UC Santa Barbara\\Santa Barbara, CA\\mhfarrell@gmail.com
  \and
  Yingjie Feng\\Tsinghua University\\Beijing, China\\fengyj@sem.tsinghua.edu.cn
}

\maketitle

\begin{abstract}
	We introduce the package \textsf{Binsreg}, which implements the binscatter methods developed by \citet*{Cattaneo-Crump-Farrell-Feng_2024_AER,Cattaneo-Crump-Farrell-Feng_2024_NonlinearBinscatter}. The package includes seven commands: \texttt{binsreg}, \texttt{binslogit}, \texttt{binsprobit}, \texttt{binsqreg}, \texttt{binstest}, \texttt{binspwc}, and \texttt{binsregselect}. The first four commands implement binscatter plotting, point estimation, and uncertainty quantification (confidence intervals and confidence bands) for least squares linear binscatter regression (\texttt{binsreg}) and for nonlinear binscatter regression (\texttt{binslogit} for Logit regression, \texttt{binsprobit} for Probit regression, and \texttt{binsqreg} for quantile regression). The next two commands focus on pointwise and uniform inference: \texttt{binstest} implements hypothesis testing procedures for parametric specifications and for nonparametric shape restrictions of the unknown regression function, while \texttt{binspwc} implements multi-group pairwise statistical comparisons. Finally, the command \texttt{binsregselect} implements data-driven number of bins selectors. The commands offer binned scatter plots, and allow for covariate adjustment, weighting, clustering, and multi-sample analysis, which is useful when studying treatment effect heterogeneity in randomized and observational studies, among many other features.\medskip
	
	\keywords{\inserttag, binscatter, binned scatter plot, nonparametrics, semiparametrics, partitioning estimators, B-splines, tuning parameter selection, confidence bands, shape and specification testing.}
\end{abstract}

\begin{center} \bigskip \today \end{center}

\newpage
%
\section{Introduction}

Data visualization is a crucial step in any statistical analysis. The classical scatter plot is a fundamental tool for visualization, used for studying how an outcome $y$ relates to a continuous covariate of interest $x$. However, in ``big data'' settings, such as the administrative data sets now common in social science and medical research, the classical scatter plot yields a dense cloud of points that is not informative. Further, there is no way to create the plot rigorously controlling for other covariates, which would be standard in any subsequent statistical analysis. A binned scatter plot, or binscatter, is a visualization method that addresses these limitations. Binscatter techniques offer flexible, yet parsimonious ways of visualizing and summarizing regression (and other) functions. Binscatters have become popular in applied microeconomics for visualization, specification testing, treatment effect heterogeneity, and other uses. See \cite{Starr-Goldfarb_2020_SMJ}, and references therein, for current usage. 

However, little was known about the statistical properties of binscatter until recently. \citet*[collectively CCFF hereafter]{Cattaneo-Crump-Farrell-Feng_2024_AER,Cattaneo-Crump-Farrell-Feng_2024_NonlinearBinscatter} provided the first foundational and comprehensive analysis of binscatter methods, including an array of theoretical and practical results that aid both in understanding current practices (i.e., validity or lack thereof) and in offering theory-based guidance for future use. Maintaining rigor and statistical uncertainty are crucial for trustworthy data visualization in scientific settings \citep{Healy2018_book,Schwabish2021_book}.

This paper introduces the package \textsf{Binsreg}, which includes seven commands implementing the main methodological results in CCFF. These commands are organized as follows.

\begin{itemize}
	\item \textit{Estimation, uncertainty quantification, and plotting}. The command \texttt{binsreg} implements canonical and extended least squares binscatter methods, while the commands \texttt{binslogit}, \texttt{binsprobit}, and \texttt{binsqreg} implement generalized nonlinear binscatter methods (i.e., Logistic regression, Probit regression and quantile regression, respectively). All commands allow for higher-order polynomial fits within bins, smoothness restrictions across bins, and covariate adjustment for estimation, uncertainty quantification and plotting, also covering higher-order derivatives and related partial effects of interest in linear and nonlinear settings, including multi-sample comparisons.
	
	\item \textit{Hypothesis testing and statistical inference}. The command \texttt{binstest} implements hypothesis testing procedures for parametric specifications and for nonparametric shape restrictions of the unknown regression function, while \texttt{binspwc} implements multi-group pairwise comparisons. These two commands offer the same flexibility and features as the estimation commands and therefore allow for linear and nonlinear binscatter methods with within-bin higher-order polynomial fits, across-bins smoothness restrictions, and semi-linear covariate adjustments, among several other features and options.
	
	\item \textit{Optimal number of bins selection}. The six commands above take as input the binning scheme to construct the binscatter approximation, which requires selecting the position of the bins as well as the total number of bins on the support of the independent variable of interest. Whenever this information is not provided, the command \texttt{binsregselect} implements data-driven selectors for the number of bins for implementation using either quantile-spaced or evenly-spaced binning/partitioning (quantile-spaced binning is chosen by default, following popular empirical practice).	
\end{itemize}

The seven commands in the package \textsf{Binsreg} offer several other important functionalities for empirical work. First, the commands incorporate, by default, mass point and degrees of freedom checks and adjustments, which improve the stability of the implementation. Second, the commands allow for multi-way fixed effects and clustering estimation and inference whenever available in the underlying statistical software platform. Third, in \texttt{Stata}, the commands offer the option of estimation and inference with multi-way fixed effects and multi-way clustering via the community-distributed package \textsf{reghdfe} \citep{Correia-Constantine_2024_reghdfe}, and also allow for using the community-distributed package \textsf{gtools} \citep{Caceres_2024_gtools}, instead of our internal implementations, to potentially increase the speed of internal computations with ultra-large datasets. Depending on the data size and structure, the commands in the package \textsf{Binsreg} may improve implementation execution speed when (i) mass point and degrees of freedom checks are turned off, (ii) the user-written packages \textsf{reghdfe} and \textsf{gtools} instead of our internal (open-source) implementations are used (in \texttt{Stata}). See Section \ref{section:Increasing Speed of Execution} for more discussion.

There exist two other user-written \texttt{Stata} commands implementing binscatter methods: \texttt{binscatter} \citep{Stepner_2017_binscatter} and \texttt{binscatter2} \citep{Droste_2019_binscatter2}. Both of those packages incorporate other covariates or fixed effects incorrectly, yielding invalid results. Even without additional controls, those two packages only give the binned scatter plot, only for piecewise constant estimation, and only in least squares regression, lacking all the theoretically-founded features of \texttt{Binsreg} such as nonlinear models, valid uncertainty visualization, formal specification and shape testing, group comparisons, and optimal binning selection. See CCFF for further discussion.

The rest of the article is organized as follows. Section \ref{section:methods} gives an overview of the main methods available in the package \textsf{Binsreg} and discusses some implementation details. Section \ref{section:illustration} gives a numerical illustration. Section \ref{section:conclusion} concludes. The software help files contain a detailed description of all available options. The latest version of the package \textsf{Binsreg} and other related materials can be found at:
\begin{center} \href{https://nppackages.github.io/binsreg/}{\texttt{https://nppackages.github.io/binsreg/}}. \end{center}

\section{Methods Overview and Implementation Details}\label{section:methods}

This section summarizes the main methods implemented in the package \textsf{Binsreg}. For further methodological and theoretical details see CCFF.

Given a random sample $(y_i,x_i,\bw'_i)$, $i=1,2,\dots,n$, where $y_i$ is a scalar response variable, $x_i$ is the scalar independent variable of interest, and $\bw_i$ is a $d$-dimensional vector of additional covariates, binscatter seeks to flexibly approximate the function
\begin{equation}\label{eq:estimand-QMLE-PE}
	\vartheta_\bw^{(v)}(x) = \frac{\partial^v}{\partial x^v} \eta(\mu_0(x)+\bw'\bgamma_0),
\end{equation}
where $\bw$ is some user-chosen evaluation point, and the underlying parameters $\mu_0(\cdot)$ and $\bgamma_0$ are defined by
\begin{equation}\label{eq:estimand-QMLE}
	(\mu_0(\cdot), \bgamma_0)=\argmin_{\mu\in\mathcal{M}, \bgamma\in\mathbb{R}^d}\; \E[\rho(y_i;\eta(\mu(x_i)+\bw_i'\bgamma))],
\end{equation}
with $\rho(\cdot;\cdot)$ and $\eta(\cdot)$ user-chosen loss and (inverse) link functions, respectively, and $\mathcal{M}$ an appropriate space of functions satisfying certain conditions. Several settings of applied interest are covered by this formulation. (For any function $f(x)$, we define $f^{(v)}(x)=d^v f(x)/dx^v$, with the usual notation $f(x)=f^{(0)}(x)$.)
\begin{itemize}
	\item \textit{Semi-linear regression}: $\rho(y;\eta)=(y-\eta)^2$ and $\eta(u)=u$. The parameter of interest becomes
	\begin{equation*}\label{eq:estimand-lsreg}
		\vartheta_\bw^{(v)}(x) = \frac{\partial^v}{\partial x^v} \E[y_i|x_i=x,\bw_i=\bw]
   	                           = \begin{cases} \mu_0(x) + \bw'\bgamma_0 & \text{if } v=0\\ \mu^{(v)}_0(x) & \text{if } v \geq 1\end{cases}.
	\end{equation*}
    For example, $\vartheta_\bw(x)$ (resp. $\vartheta^{(1)}_\bw(x)$) corresponds to the average (partial) effect of $x$ on $y$ for level $\bw_i=\bw$. In this setting, $\vartheta^{(v)}_{\mathbf{0}}(x)=\mu^{(v)}_0(x)$, which may be of interest in some applications.

	\item \textit{Logistic/Probit regression}: $\rho(y;\eta)=-y\log \eta-(1-y)\log (1-\eta)$ and $\eta(u)$ denotes the (inverse) link function of Logistic or Probit regression. The parameter of interest becomes
	\begin{equation*}\label{eq:estimand-binaryreg}
		\vartheta_\bw^{(v)}(x) = \frac{\partial^v}{\partial x^v} \E[y_i|x_i=x,\bw_i=\bw]
		                       = \frac{\partial^v}{\partial x^v} \eta(\mu_0(x) + \bw'\bgamma_0),
	\end{equation*}
	which coincides with the usual average (partial) effect in binary response models.
	
	\item \textit{Quantile regression}: $\rho(y;\eta)= \ell_\tau(y-\eta)$ and $\eta(u)=u$, where $\ell_\tau(u)$ denotes the check function associated with the $\tau$-th quantile. The parameter of interest becomes
	\begin{equation*}\label{eq:estimand-qreg}
		\vartheta_\bw^{(v)}(x) = \frac{\partial^v}{\partial x^v} Q_\tau(y_i|x_i=x,\bw_i=\bw),
	\end{equation*}
	where $Q_\tau(y_i|x_i=x,\bw_i=\bw)=\mu_0(x) + \bw'\bgamma_0$ denotes the conditional $\tau$-th quantile regression function of $y_i$ given $x_i=x,\bw_i=\bw$.
\end{itemize} 

The different parameters above, as well as many others, are determined by the choice of loss function $\rho(\cdot;\cdot)$ and (inverse) link function $\eta(\cdot)$. In the above formulation, we assume that the models are correctly specified relative to the true data generating process (i.e., relative to the assumptions on the probability distribution of the data $(y_i,x_i,\bw'_i)$, $i=1,2,\dots,n$), which is also assumed to be an i.i.d. sample. However, in many settings, the choices of $\rho(\cdot;\cdot)$ and $\eta(\cdot)$ are only working models, which may not lead to the underlying target parameter but rather only to an approximation thereof in a principled way. That is, under incorrect specification, the parameter $\vartheta_\bw^{(v)}(x)$ can only be interpreted as the solution to the minimization in \eqref{eq:estimand-QMLE}. Under dependent data, binscatter can still be applied but statistical guarantees for parameter estimation and for uncertainty quantification are not available.

\subsection{Binscatter Construction}

To approximate $\mu_0(x)$ and its derivatives in model \eqref{eq:estimand-QMLE}, binscatter first partitions the support of $x_i$ into $J$ quantile-spaced bins, leading to the partitioning scheme:
\[\widehat{\Delta} = \{\widehat{\mathcal{B}}_1, \dots, \widehat{\mathcal{B}}_J\},\qquad
\widehat{\mathcal{B}}_j = \begin{cases}
\big[x_{(1)}, x_{(\lfloor n/J \rfloor)}\big)                            & \quad \text{if } j=1,\\
\big[x_{(\lfloor n(j-1)/J \rfloor)}  , x_{(\lfloor nj/J \rfloor)}\big)  & \quad \text{if } j=2,\dots,J-1\\
\big[x_{(\lfloor n(J-1)/J \rfloor)}, x_{(n)}\big]                       & \quad \text{if } j=J,
\end{cases},
\]
where $x_{(i)}$ denotes the $i$-th order statistic of the sample $\{x_1,x_2,\dots,x_n\}$, $\lfloor \cdot \rfloor$ is the floor operator, and $J<n$. Each estimated bin $\widehat{\mathcal{B}}_j$ contains roughly the same number of observations $N_j=\sum_{i=1}^n \I_{\widehat{\mathcal{B}}_j}(x_i)$, where $\I_\mathcal{A}(x)=\I(x\in \mathcal{A})$ with $\I(\cdot)$ denoting the indicator function. This binning approach is the most popular in empirical work but, for completeness, all commands in the package \textsf{Binsreg} also allow for evenly-spaced binning and user-specified binning. See below for more implementation details.

Given the quantile-spaced partitioning/binning scheme, for a choice of number of bins $J$, and a choice of loss function $\rho(\cdot;\cdot)$ and (inverse) link function $\eta(\cdot)$, the generalized nonlinear binscatter estimator of the $v$-th derivative $\vartheta_\bw^{(v)}(x)$ of $\eta(\mu_0(x)+\bw'\bgamma_0)$ in \eqref{eq:estimand-QMLE-PE}, employing a $p$-th order polynomial approximation within each bin, imposing $(s-1)$-times differentiability across bins, and adjusting for additional covariates $\bw_i$, is
\begin{equation}\label{eq:estimator-QMLE-PE}
	\widehat{\vartheta}_\bw^{(v)}(x) = \frac{\partial^v}{\partial x^v} \eta(\widehat{\mu}(x)+\bw'\widehat{\bgamma})
\end{equation}
where
\begin{equation}\label{eq:estimator-QMLE}
	\widehat{\mu}^{(v)}(x) = \widehat{\bb}_{p,s}^{(v)}(x)'\widehat{\bbeta}, \qquad
	\begin{bmatrix}\;\widehat{\bbeta}\;\\\;\widehat{\bgamma}\;\end{bmatrix}
	= \argmin_{\bbeta,\bgamma} \sum_{i=1}^{n} \rho\Big(y_i; \;\eta\big(\widehat{\bb}_{p,s}(x_i)'\bbeta+\bw_{i}'\bgamma\big)\Big),
\end{equation}
with $s \leq p$, $v \leq p$ and $\widehat{\bb}_{p,s}(x) = \widehat{\mathbf{T}}_s \widehat{\bb}_{p,0}(x)$ with
\[\widehat{\bb}_{p,0}(x)
=\begin{bmatrix}\;\I_{\widehat{\mathcal{B}}_1}(x) & \I_{\widehat{\mathcal{B}}_2}(x)& \cdots & \I_{\widehat{\mathcal{B}}_J}(x) \;\end{bmatrix}'
\otimes
\begin{bmatrix}\; 1 & x & \cdots & x^p \;\end{bmatrix}',
\]
being the $p$-th order polynomial basis of approximation within each bin, hence of dimension $(p+1)J$, and $\widehat{\mathbf{T}}_s$ being a $[(p+1)J-(J-1)s]\times (p+1)J$ matrix of linear restrictions ensuring that the $(s-1)$-th derivative of $\widehat{\mu}(x)$ is continuous.

When $s=0$, $\widehat{\mathbf{T}}_0=\mathbf{I}_{(p+1)J}$, the identity matrix of dimension $(p+1)J$, and therefore no restrictions are imposed: $\widehat{\bb}_{p,0}(x)$ is the basis used for (disjoint) piecewise $p$-th order polynomial fits. Consequently, the binscatter $\widehat{\mu}(x)$ is discontinuous at the bins' edges whenever $s=0$. On the other hand, $p\geq s$ implies that a $p$-th order polynomial fit is constructed within each bin $\widehat{\mathcal{B}}_j$, in which case setting $s=1$ forces these fits to be connected at the boundaries of adjacent bins (leading to a continuous but nondifferentiable function), $s=2$ forces these fits to be connected and continuously differentiable at the boundaries of adjacent bins, and so on. Enforcing smoothness on binscatter boils down to incorporating restrictions on the basis of approximation. The resulting constrained basis, $\widehat{\bb}_{p,s}(x)$, corresponds to a choice of spline basis for approximation of $\mu_0(\cdot)$ in \eqref{eq:estimand-QMLE}, with estimated quantile-spaced knots according to the partition $\widehat{\Delta}$. The package \textsf{Binsreg} employs $\widehat{\mathbf{T}}_s$ leading to B-splines, which tend to have very good finite sample properties.

The binscatter estimator $\widehat{\vartheta}_\bw^{(v)}(x)$ in \eqref{eq:estimator-QMLE-PE} is a plug-in estimator for \eqref{eq:estimand-QMLE}. Returning to the settings of applied interest mentioned previously, we have:
\begin{itemize}
	\item \textit{Semi-linear regression}: $\rho(y;\eta)=(y-\eta)^2$ and $\eta(u)=u$. This case corresponds to linear least squares binscatter, where the estimator becomes
	\begin{equation*}\label{eq:estimator-lsreg}
		\widehat{\vartheta}_\bw^{(v)}(x) = \frac{\partial^v}{\partial x^v} \widehat{\E}[y_i|x_i=x,\bw_i=\bw]
		= \begin{cases} \widehat{\mu}(x) + \bw'\widehat{\bgamma} & \text{if } v=0\\ \widehat{\mu}^{(v)}(x) & \text{if } v \geq 1\end{cases}.
	\end{equation*}
	This estimator is obtained by running the linear least squares regression of $y_i$ on $(\widehat{\bb}_{p,s}(x_i)',\bw_i')$, and then constructing predicted values at $(x,\bw')$ for $v=0$, or predicted values $\widehat{\mu}^{(v)}(x) =\widehat{\bb}_{p,s}^{(v)}(x)'\widehat{\bbeta}$ for $v \geq 1$. The command \texttt{binsreg} provides implementation for this case.
		
	\item \textit{Logistic/Probit regression}: $\rho(y;\eta)=-y\log \eta-(1-y)\log (1-\eta)$ and $\eta(u)$ denotes the (inverse) link function of Logistic or Probit regression. This case corresponds to nonlinear Logistic or Probit binscatter, where the estimator becomes
	\begin{equation*}\label{eq:estimator-binaryreg}
		\widehat{\vartheta}_\bw^{(v)}(x) = \frac{\partial^v}{\partial x^v} \widehat{\E}[y_i|x_i=x,\bw_i=\bw]
		= \frac{\partial^v}{\partial x^v} \eta(\widehat{\mu}(x) + \bw'\widehat{\bgamma}).
	\end{equation*}
	This estimator is obtained by running the Logit or Probit nonlinear regression of $y_i$ on $(\widehat{\bb}_{p,s}(x_i)',\bw_i')$, and constructing predicted values at $(x,\bw')$ for $v=0$, or derivatives thereof for $v \geq 1$. The commands \texttt{binslogit} and \texttt{binsprobit} provide implementation for these cases. These two commands only allow $v=0$ or $1$.
	
	\item \textit{Quantile regression}: $\rho(y;\eta)= \ell_\tau(y-\eta)$ and $\eta(u)=u$, where $\ell_\tau(u)$ denotes the check function associated with the $\tau$-th quantile. This case corresponds to nonlinear, non-differentiable quantile regression binscatter, where the estimator becomes
	\begin{equation*}\label{eq:estimator-qreg}
		\widehat{\vartheta}_\bw^{(v)}(x) = \frac{\partial^v}{\partial x^v} \widehat{Q}_\tau(y_i|x_i=x,\bw_i=\bw),
	\end{equation*}
	where $\widehat{Q}_\tau(y_i|x_i=x,\bw_i=\bw)=\widehat{\mu}(x) + \bw'\widehat{\bgamma}$ denotes the estimate of the conditional $\tau$-th quantile  function of $y_i$ given $(x_i,\bw_i')$. This estimator is obtained by running the quantile regression of $y_i$ on $(\widehat{\bb}_{p,s}(x_i)',\bw_i')$, and constructing predicted values at $(x,\bw')$ for $v=0$, or derivatives thereof for $v \geq 1$. The command \texttt{binsqreg} provides implementation for this case.
\end{itemize}

In practice, the binscatter estimator $\widehat{\vartheta}_\bw^{(v)}(x)$ needs to be evaluated at some point $\bw$. Typical choices are $\bw=\mathbf{0}$, $\bw=\bar{\bw}=\frac{1}{n}\sum_{i=1}^n \bw_i$, or $\bw=\text{median}(\bw_i)$, with $\mathbf{0}$ denoting a vector of zeros and $\text{median}(\bw_i)$ denoting the empirical median of each component in $\bw_i$. For discrete variables in $\bw$ it is natural to set those components to some base category (e.g., zero for binary variables), while for continuous variables it may be better to set those components at some other value (e.g., mean or some quantile). The point of evaluation affects the visual and statistical properties of the binscatter, as discussed below.

\subsubsection{Canonical Binscatter}

Canonical binscatter, as implemented in the packages \texttt{binscatter} and \texttt{binscatter2}, corresponds to linear least squares regression ($\rho(y;\eta)=(y-\eta)^2$ and $\eta(u)=u$) with $p=s=0$ and without covariate adjustment (i.e., not including $\bw_i$ in \eqref{eq:estimator-QMLE}). Specifically, in canonical binscatter the basis $\widehat{\bb}_{0,0}(x)$ is a $J$-dimensional vector of orthogonal dummy variables, that is, the $j$-th component of $\widehat{\bb}_{0,0}(x)$ records whether the evaluation point $x$ belongs to the $j$-th bin in the partition $\widehat{\Delta}$. Therefore, canonical binscatter can be expressed as the collection of $J$ sample averages of the response variable $y_i$, one for each bin: $\bar{y}_j = \frac{1}{N_j} \sum_{i=1}^n \I_{\widehat{\mathcal{B}}_j}(x_i) y_i$ for $j=1,2,\dots,J$. Empirical work employing canonical binscatter typically plots these binned sample averages along with some other estimate(s) of the regression function $\mu_0(x)$.

\subsubsection{Covariate-Adjusted Binscatter}

Prior work employing binscatter methods, including the packages \texttt{binscatter} and \texttt{binscatter2}, not only considered exclusively least squares regressions with $p=s=0$, but also performed covariate adjustment by residualization. To be precise, first the residuals from the linear regressions of $y_i$ on $(1,\bw_i')$ and of $x_i$ on $(1,\bw_i')$ were computed, and then a canonical binscatter was estimated using those residuals. We call this approach residualized canonical binscatter.

CCFF showed that residualized canonical binscatter is very hard to rationalize or justify, and will lead to an inconsistent estimator of $\vartheta_\bw^{(v)}(x)$ unless very special assumptions hold, even when the statistical model is correctly specified. In contrast, our proposed approach for covariate adjustment \eqref{eq:estimator-QMLE} is justified via the model \eqref{eq:estimand-QMLE}, and is therefore principled and interpretable. Even when model \eqref{eq:estimand-QMLE} is misspecified, the approach to covariate adjustment employed by the package \textsf{Binsreg} enjoys a natural probability limit interpretation, while the residualization approach does not. See CCFF for more discussion, numerical examples, and technical details.

\subsubsection{Main implementation details}

The four estimation commands \texttt{binsreg}, \texttt{binslogit}, \texttt{binsprobit}, and \texttt{binsqreg} implement, respectively, least squares, Logit, Probit, and quantile regression binscatter estimators for a given choice of partitioning/binning $\widehat{\Delta}$. The option \texttt{deriv()} is used to set the value of $v$ and the option \texttt{at()} is used to set the value of $\bw$ in the estimator $\widehat{\vartheta}_\bw^{(v)}(x)$. The options \texttt{dots(p s)} and \texttt{line(p s)} generate ``dots'' and a ``line'' tracing out two distinct implementations of $\widehat{\vartheta}_\bw^{(v)}(x)$ with the corresponding choices of $p$ and $s$ selected in each case, but using the same values of $v$ and $\bw$. If \texttt{dots(T)} (or \texttt{line(T)}) is specified, \texttt{dots(0 0)} (or \texttt{line(0 0)}) is used unless the degree $p$ or smoothness $s$ selection is requested via the option \texttt{pselect()} or \texttt{sselect()} (see details in the next subsection).

The defaults are to estimate the level of the function ($v=0$, \texttt{deriv(0)}) and evaluate the covariates at the mean ($\bw_i=\bar{\bw}$, \texttt{at(mean)}). Evaluating the covariates at different points can effect a level shift of the plotted point estimates and change the statistical uncertainty, such as confidence bands (see the supplemental appendices of CCFF). The option \texttt{at()} allows for the mean, median, and a vector of zeros, the last being useful for dummy variables or fixed effects. For high-dimensional fixed effects, the user-contributed \texttt{reghdfe} package can be called with the \texttt{absorb} option. In this case the fixed effects are normalized within \texttt{reghdfe} and the point of evaluation cannot be set.

For example, when using \texttt{binsreg}, the default implementation yields the estimate $\widehat{\vartheta}_{\bar{\bw}}(x) = \widehat{\E}[y_i|x_i=x,\bw_i=\bar{\bw}]= \widehat{\mu}(x) + \bar{\bw}'\widehat{\bgamma}$. Thus, \texttt{dots(0 0)} leads to ``dots'' representing sample averages within each bin for the ``long'' regression with $\bw_i=\bar{\bw}$. In particular, if $\bw_i$ are not included, then the default coincides with Canonical Binscatter (i.e., the same results would be obtained using the packages \texttt{binscatter} and \texttt{binscatter2} for the same $J$). The line option is muted by default, and needs to be set explicitly to appear in the resulting plot: for example, the option \texttt{line(3 3)} adds a line tracing out $\widehat{\mu}^{(v)}(x)$, implemented with $p=3$ and $s=3$, a cubic B-spline approximation of $\mu_0^{(v)}(x)$. 

The common partitioning/binning used by the four estimation commands across all implementations is set to be quantile-spaced for some choice of $J$. The option \texttt{nbins()} sets $J$ manually (e.g., \texttt{nbins(20)} corresponds to $J=20$ quantile-spaced bins), but if this option is not supplied then the companion command \texttt{binsregselect} is used to choose $J$ in a fully data-driven way, as described below. As an alternative, an evenly-spaced or user-specified partitioning/binning can be implemented via the option \texttt{binspos()}.

Several other options are available for the four estimation commands, including multi-way fixed effects and multi-way clustering adjustments. Each command has an accompanying help file with complete details.

\subsection{Choosing the Number of Bins}

From a statistical point of view, $J$ is the main tuning parameter of a binscatter and, as usual for nonparametrics, one must assume $J \to \infty$ for consistent estimation. To provide an optimal, data-driven choice of $J$, CCFF developed valid integrated mean squared error (IMSE) approximations for generalized nonlinear binscatter in the context of model \eqref{eq:estimand-QMLE}. These expansions give IMSE-optimal selection of the number bins $J$, depending on polynomial order $p$ within bins and smoothness level $s$ across bins, the target estimand set by the derivative order $v$, and the evaluation point of interest $\bw$ for covariate adjustment. Specifically, the IMSE-optimal choice of $J$ is
\[J_\texttt{IMSE} = \left\lceil \left(\frac{2(p-v+1)\mathscr{B}_n(p,s,v)}{(1+2v)\mathscr{V}_n(p,s,v)}
\right)^{\frac{1}{2p+3}} \; n^{\frac{1}{2p+3}} \right\rceil,\]
where $\lceil \cdot \rceil$ denotes the ceiling operator, $\mathscr{B}_n(p,s,v)$ and $\mathscr{V}_n(p,s,v)$ represent an approximation to the integrated (squared) bias and variance of $\widehat{\vartheta}_\bw^{(v)}(x)$ respectively, and the three integer choices must respect $p\geq s\geq0$ and $p\geq v\geq0$. The constants $\mathscr{B}_n(p,s,v)$ and $\mathscr{V}_n(p,s,v)$ depend on the partitioning scheme and binscatter estimator used. Note that the package \texttt{Binsreg} allows for the standard option \texttt{vce()} to specify variance-covariance estimation methods, which can also affect the variance constant $\mathscr{V}_n(p,s,v)$. 

For simplicity, the command \texttt{binsregselect} in the package \textsf{Binsreg} implements number of bins selection based on the IMSE expansion for the linear least squares binscatter. For generalized nonlinear binscatter (Logistic, Probit, or quantile regression), the number of bins $J$ given by the command \texttt{binsregselect} still has the ``correct'' rate (the same order as that of the IMSE-optimal one). Thus, confidence bands and testing procedures  based on such choices of $J$ and the robust bias correction strategy described below are still valid even in the general nonlinear case. 

Both IMSE constants, $\mathscr{B}_n(p,s,v)$ and $\mathscr{V}_n(p,s,v)$, can be estimated consistently using a preliminary choice of $J$. Thus, our implementation offers two $J$ selectors.
\begin{itemize}
	\item $\widehat{J}_\mathtt{ROT}$: implements a rule-of-thumb (ROT) approximation for the constants  $\mathscr{B}_n(p,s,v)$ and $\mathscr{V}_n(p,s,v)$, employing a trimmed-from-below Gaussian reference model for the density of $x_i$, and global polynomial approximations for the other two unknown features needed, $\mu_0^{(v)}(x)$ and $\V[y_i|x_i=x, \bw_i=\bw]$. This $J$ selector employs the correct rate but an inconsistent constant approximation.
	
	\item $\widehat{J}_\mathtt{DPI}$: implements a direct-plug-in (DPI) approximation for the constants $\mathscr{B}_n(p,s,v)$ and $\mathscr{V}_n(p,s,v)$, based on the desired binscatter, set by the choices $p$ and $s$, and employing a preliminary $J$. If a preliminary $J$ is not provided by the user, then $J=\max\{\widehat{J}_\mathtt{ROT},\lceil (\frac{2(p-v+1)}{1+2v}n)^{\frac{1}{2p+3}}\rceil\}$ is used for DPI implementation. This $J$ selector employs the correct rate as well as consistent estimators of the appropriate constants. Default implementation uses $\widehat{J}_\mathtt{DPI}$ whenever $J$ is not specified.
\end{itemize}

Implementing a binscatter with $J=J_\texttt{IMSE}$ is optimal from a statistical point of view (for valid estimation, testing, and uncertainty quantification), but sometimes a fixed, user-chosen number of bins, denoted by $J=\mathtt{J}$, may yield a more visually appealing binscatter (albeit with the caveat that the implied estimator may be inaccurate). Many applications in the past used round numbers such as 10, 20, 50, or 100 bins. Further, a fixed $J=\mathtt{J}$ can also be directly interpretable as a discretized version of $x$. For instance, setting $J=\mathtt{J} = 10$ yields a binscatter that allows for comparison across deciles of $x$, for example comparing those in the top decile of earnings to those at the bottom. Setting $\mathtt{J}= 100$ can be used to compare across percentiles, and is commonly used when studying ranks.

To balance the potential for an appealing visualization based on a fixed $J = \mathtt{J}$ with the desire for statistical validity, CCFF developed a novel method for selecting the polynomial order $p$ (and along with it, the smoothness $s$) as a function of the fixed  ${\tt J}$, as opposed to $J_\texttt{IMSE} = J_\texttt{IMSE}(p,s,v)$ which does the reverse. Specifically, CCFF proposed to look for the values $p^*$ and $s^*$ (in a pre-specified range) such that 
$J_\texttt{IMSE}(p^*,s^*,v)$ is approximately equal to the researcher's chosen $\mathtt{J}$. That is, finding the $p$ and $s$ for which the chosen $\mathtt{J}$ would be IMSE-optimal. Although $p \to \infty$ is not allowed in the theory of CCFF, and so this choice is somewhat ad-hoc, it may effectively reduce the bias of the binscatter. 

\subsubsection{Main implementation details}

Unless $J$ is manually specified, all commands in the \texttt{Binsreg} package employ
the command \texttt{binsregselect} to implement ROT and DPI data-driven, IMSE-optimal selection of $J$ for all possible choices of $p\geq v,s \geq0$, and for both quantile-spaced or evenly-spaced partitioning/binning. For DPI implementation, the user can provide the initialization value of $J$ via the option \texttt{nbinsrot()} or, if not provided, then $\widehat{J}_\mathtt{ROT}$ is used. 

As discussed above, instead of selecting the number of bins $J$, an alternative strategy is setting a fixed value for $J$ and implementing (ROT or DPI) data-driven, IMSE-optimal selection of the degree of polynomial $p$ and/or the number of smoothness constraints $s$. The command \texttt{binsregselect} implements this selection procedure if (i) the number of bins $J$ is supplied via the option \texttt{nbins()} and (ii) a range for searching for the optimal $p$ or $s$ is supplied via the option \texttt{pselect()} or \texttt{sselect()}.

Several other options are available for the command \texttt{binsregselect}, including the possibility of generating an output file with the IMSE-optimal partitioning/binning structure selected and the corresponding grid of evaluation points, which can be used by the other six companion commands for plotting, simulation, testing, and other calculations.

\subsection{Confidence Intervals}

Both confidence intervals and confidence bands for the unknown function $\vartheta_\bw^{(v)}(x)$ are constructed employing the same type of Studentized $t$-statistic:
\[T_p(x) = \frac{\widehat{\vartheta}_\bw^{(v)}(x) - \vartheta_\bw^{(v)}(x)}{\sqrt{\widehat{\Omega}(x)/n}}, \qquad 0\leq v,s\leq p,\]
where the binscatter variance estimator is of the usual ``sandwich'' form
\[\widehat{\Omega}(x) = \widehat{\bb}_{p,s}^{(v)}(x)'\widehat{\bQ}^{-1}\widehat{\bSigma}\widehat{\bQ}^{-1}\widehat{\bb}_{p,s}^{(v)}(x)(\eta^{(1)}(\widehat{\mu}(x)+\bw'\widehat{\bgamma}))^2,\]
\[\widehat{\bQ} = \frac{1}{n}\sum_{i=1}^n \widehat{\bb}_{p,s}(x_i)\widehat{\bb}_{p,s}(x_i)'\widehat{\Psi}_{i,1}\widehat{\eta}_{i,1}^2,\]
\[\widehat{\bSigma} = \frac{1}{n}\sum_{i=1}^n \widehat{\bb}_{p,s}(x_i)\widehat{\bb}_{p,s}(x_i)' \widehat{\eta}_{i,1}^2 \psi(y_i, \widehat{\eta}_{i,0})^2\]
with $\widehat{\Psi}_{i,1}$ a consistent estimator of $\Psi_{i,1}=\frac{\partial}{\partial \eta}\E[\psi(y_i,\eta)|x_i,\bw_i]\big|_{\eta=\eta_{i,0}}$, $\widehat{\eta}_{i,v}$ a consistent estimator of $\eta_{i,v}=\eta^{(v)}(\mu_0(x_i)+\bw_i'\bgamma_0)$ for $v=0,1$, and $\psi(y,u)$ the (weak) derivative of $\rho(\cdot;u)$ with respect to $u$. See CCFF for details and omitted formulas. In practice, these estimators are implemented using the base commands from the statistical software.

CCFF showed that $T_p(x) \to_d \mathsf{N}(0,1)$ pointwise in $x$, that is, for each evaluation point $x$ on the support of $x_i$, provided the misspecification error introduced by binscatter is removed from the distributional approximation. Such a result justifies asymptotically valid confidence intervals for $\vartheta_\bw^{(v)}(x)$, pointwise in $x$, after bias correction. Specifically, for each $x$, the $(1-\alpha)\%$ confidence interval takes the form:
\[\widehat{I}_p(x) = \Big[ \; \widehat{\vartheta}_\bw^{(v)}(x) \pm \Phi^{-1}(1-\alpha/2) \cdot \sqrt{\widehat{\Omega}(x)/n} \; \Big],
\qquad 0\leq v,s\leq p,
\]
where $\Phi(u)$ denotes the distribution function of a standard normal random variable (e.g., $\Phi^{-1}(1-0.05/2) \approx 1.96$ for a $95\%$ Gaussian confidence intervals), and provided the choice of $J$ is such that the misspecification error can be ignored.

However, employing an IMSE-optimal binscatter (i.e., setting $J=J_\texttt{IMSE}$ for the selected polynomial order $p$) introduces a first-order misspecification error leading to invalidity of these confidence intervals, and hence cannot be directly used to form the confidence intervals $\widehat{I}_p(x)$ in general. To address this problem, we rely on a simple application of robust bias correction \citep{Calonico-Cattaneo-Titiunik_2014_ECMA,Calonico-Cattaneo-Farrell_2018_JASA,Cattaneo-Farrell-Feng_2020_AoS,Calonico-Cattaneo-Farrell_2022_Bernoulli} to form valid confidence intervals based on IMSE-optimal binscatter, that is, without altering the partitioning scheme $\widehat{\Delta}$ used.

Our implementation employs robust bias-corrected binscatter confidence intervals as follows. First, for a given choice of $p$, select the number of bins in $\widehat{\Delta}$ according to $J=J_\mathtt{IMSE}$, which gives an IMSE-optimal binscatter (point estimator). Then, employ the confidence interval $\widehat{I}_{p+q}(x)$ with $q\geq1$, which gives a valid confidence interval: $\P\big[ \vartheta_\bw^{(v)}(x) \in \widehat{I}_{p+q}(x) \big] \to 1 -\alpha$, for all $x$.

The same strategy is applied when the degree/smoothness selection described previously is implemented. First, for a fixed choice of $\mathtt{J}$, select the ``optimal'' degree of polynomial $p$  (i.e., the resulting $J_{\mathtt{IMSE}}$ is the closest to the chosen $\mathtt{J}$). Then, increase the degree of polynomial to construct the confidence intervals $\widehat{I}_{p+q}(x)$. This idea is also employed in the construction of confidence bands and hypothesis testing procedures and can be fully implemented using the options \texttt{pselect()} and \texttt{sselect()}. Since the degree/smoothness selection is \textit{not} the default in the package \texttt{Binsreg}, our discussion focuses on robust bias correction based on IMSE-optimal $J$ selection.

\subsubsection{Main implementation details}

The four estimation commands \texttt{binsreg}, \texttt{binslogit}, \texttt{binsprobit}, and \texttt{binsqreg} implement confidence intervals, and report them as part of the final binned scatter plot. Specifically, the option \texttt{ci(p s)} estimates confidence intervals with the corresponding choices of $p$ and $s$ selected, and plots them as vertical segments along the support of $x_i$. If \texttt{ci(T)} is specified, \texttt{ci(1 1)} is used unless the degree $p$ or smoothness $s$ selection is requested via the option \texttt{pselect()} or \texttt{sselect()} as described before. The confidence interval option is muted by default, and needs to be set explicitly to appear in the resulting plot. The implementation is done over a grid of evaluation points, which can be modified via the option \texttt{cigrid()}, and the desired level is set by the option \texttt{level()}. Notice that \texttt{dots(p s)}, \texttt{lines(p s)}, and \texttt{ci(p s)} may all take different choices of $p$ and $s$, which allows for robust bias correction implementation of the confidence intervals and permits incorporating different levels of smoothness restrictions.

\subsection{Confidence Bands}

In many empirical applications of binscatter, the goal is to conduct inference about the entire function $\vartheta_\bw^{(v)}(x)$, simultaneously, that is, uniformly over all $x$ values of the support of $x_i$. This goal is fundamentally different from pointwise inference. A leading example of uniform inference is reporting confidence bands for $\vartheta_\bw(x)$ and its derivatives, which are different from (pointwise) confidence intervals. The package \textsf{Binsreg} offers asymptotically valid constructions of both confidence intervals, as discussed above, and confidence bands, which can be implemented with the same choices of $(p,s)$ used to construct $\widehat{\vartheta}_\bw^{(v)}(x)$ or different ones.

Following the theoretical work in CCFF, for a choice of $p$ and partition/binning of size $J$, the $(1-\alpha)\%$ confidence band for $\vartheta_\bw^{(v)}(x)$ is:
\[\widehat{I}_p(\cdot) = \Big[ \; \widehat{\vartheta}_\bw^{(v)}(\cdot) \pm \cval \cdot \sqrt{\widehat{\Omega}(\cdot)/n} \; \Big],
\qquad 0\leq v,s\leq p,
\]
where the quantile value $\cval$ is now approximated via simulations using
\[\cval = \inf\Big\{c\in\mathbb{R}_+:\P\Big[ \sup_{x} \big|\widehat{Z}_{p}(x)\big|\leq c \;\Big| \;\bD \Big]\geq 1-\alpha \Big\},\]
with $\bD= ((y_i,x_i,\bw_i'):1\leq i\leq n)$ denoting the original data,
\[\widehat{Z}_p(x) = \frac{\widehat{\bb}_{p,s}^{(v)}(x)'\widehat{\bQ}^{-1}\widehat{\bSigma}^{-1/2}}{\sqrt{\widehat{\Omega}(x)/n}}\bN_{K}^\star,
  \quad K=(p+1)J-(J-1)s,
  \quad 0\leq v,s\leq p,
\]
and $\bN_K^\star \thicksim \mathsf{N}(\mathbf{0},\mathbf{I})$ being a $K$-dimensional standard normal random vector independent of the data $\bD$. The distribution of $\sup_{x} \big|T_{p}(x)\big|$, which is unknown, is approximated by that of $\sup_{x} \big|\widehat{Z}_{p}(x)\big|$ conditional on the data $\bD$, which can be simulated by taking repeated samples from $\bN_{K}^\star$ and recomputing the supremum each time. In other words, the quantiles used to construct confidence bands can be approximated by resampling from the standard normal random vector $\bN_K^\star$, keeping fixed the data $\bD$ (and hence all quantities depending on it). See CCFF for more details.

A confidence band covers the entire function, $\vartheta_\bw^{(v)}(x)$, $(1-\alpha)\%$ of the time in repeated sampling, whenever the misspecification error can be ignored. As before, we recommend employing robust bias correction to remove misspecification error introduced by binscatter, that is, following the same logic discussed above for the case of confidence interval construction. To be more precise, first $p$ is chosen, along with $s$ and $v$, and the optimal partitioning/binning is selected according to $J=J_\mathtt{IMSE}$. Then, valid confidence bands are constructed using $\widehat{I}_{p+q}(x)$ with $q\geq1$: $\P\big[ \vartheta_\bw^{(v)}(x) \in \widehat{I}_{p+q}(x), \; \text{for all } x \big] \to 1 -\alpha$.

Moreover, the visual appearance of the confidence band will be impacted by the chosen evaluation point $\bw$, and the researcher needs to be careful when using the band as visual aids in parametric specification testing. See CCFF, and in particular Section SA-1.2 of \cite{Cattaneo-Crump-Farrell-Feng_2024_AER}.

\subsubsection{Main implementation details}

The four estimation commands \texttt{binsreg}, \texttt{binslogit}, \texttt{binsprobit}, and \texttt{binsqreg} implement confidence bands, and report them as part of the final binned scatter plot. The option \texttt{cb(p s)} estimates an asymptotically valid confidence band with the corresponding choices of $p$ and $s$ selected, and plots it as a shaded region along the support of $x_i$. If \texttt{cb(T)} is specified, \texttt{cb(1 1)} is used unless the degree $p$ or smoothness $s$ selection is requested via the option \texttt{pselect()} or \texttt{sselect()} as described before. The confidence band option is muted by default, and needs to be set explicitly to appear in the resulting plot. The implementation is done over a grid of evaluation points, which can be modified via the option \texttt{cbgrid()}, and the desired level is set by the option \texttt{level()}. The options \texttt{dots(p s)}, \texttt{lines(p s)}, \texttt{ci(p s)}, and \texttt{cb(p s)} can all take different choices of $p$ and $s$, which allows for robust bias correction implementations, as well as many other practically relevant possibilities.

\subsection{Parametric Specification Testing}

In addition to implementing binscatter and producing binned scatter plots, with both point estimation and uncertainty visualization, the package \textsf{Binsreg} also allows for formal testing of substantive hypotheses. The command \texttt{binstest} implements two types of substantive hypothesis tests about $\vartheta_\bw^{(v)}(x)$: (i) parametric specification testing and (ii) nonparametric shape restriction testing. This subsection discusses the former, while the next subsection discusses the latter.

For a choice of $(p,s,v)$, and partitioning/binning scheme of size $J$, the implemented parametric specification testing approach contrasts a (nonparametric) binscatter approximation $\widehat{\vartheta}_\bw^{(v)}(x)$ of $\vartheta_\bw^{(v)}(x)$ with a hypothesized parametric specification of the form $\vartheta_\bw(x)=\eta(M_\bw(x;\btheta,\bgamma_0))$ where $M_\bw(x;\btheta,\bgamma_0)=m(x;\btheta)+\bw'\bgamma_0$ for some $m(\cdot)$ known up to a finite parameter $\btheta$, which can be estimated using the available data. Formally, the null and alternative hypotheses are, respectively,
\begin{align*}
\dot{\mathsf{H}}_0&:\quad \sup_{x} \Big|\vartheta_\bw^{(v)}(x) - \frac{\partial^v}{\partial x^v}\eta(M_\bw(x;\btheta,\bgamma_0))\Big|=0, \quad \text{ for some } \btheta, \qquad vs.\\
\dot{\mathsf{H}}_\text{A}&:\quad \sup_{x} \Big|\vartheta_\bw^{(v)}(x) - \frac{\partial^v}{\partial x^v}\eta(M_\bw(x;\btheta,\bgamma_0))\Big|>0, \quad \text{ for all } \btheta,
\end{align*}
for a choice of $v$.

For example, excluding additional covariates $\bw_i$, $\widehat{\mu}(x)$ is compared to $\bar{y}=\frac{1}{n}\sum_{i=1}^n y_i$ in order to assess whether there is a relationship between $y_i$ and $x_i$ or, more formally, whether $\mu_0(x)$ is a constant function. Similarly, it is possible to formally test for a linear, quadratic, or even nonlinear parametric relationship $\mu_0(x)=m(x,\btheta)$, where $\btheta$ would be estimated from the data under the null hypothesis, that is, assuming that the postulated relationship is indeed correct.

However, when additional covariates $\bw_i$ are included, CCFF showed that the special case of the test $\dot{\mathsf{H}}_0\; vs. \;\dot{\mathsf{H}}_{\text{A}}$ with $v=0$ could give conclusions that are sensitive to the user-selected point of evaluation $\bw$, which implies that the common practice of visually examining a binned scatter plot compared to a parametric specific could be misleading. See Section SA-1.2 of \cite{Cattaneo-Crump-Farrell-Feng_2024_AER} for further details.

To avoid this issue, and motivated by the fact that the central point of binscatter is to study how $y_i$ relates to $x_i$, controlling for $\bw_i$, we advocate reformulating the hypothesis as pertaining to the \textit{derivative} of $\mu_0(x)$, instead of the level. For example, to test if $\mu_0(x)$ is linear, one can test if it has a constant first derivative, i.e.,  $\mathsf{H}_0: \sup_{x}|\mu_0^{(1)}(x)-a|=0$ for some $a$, $vs.$ $\mathsf{H}_{\text{A}}: \sup_{x}|\mu_0^{(1)}(x)-a|>0$. This can be viewed as a special case of the test $\dot{\mathsf{H}}_0\; vs. \;\dot{\mathsf{H}}_{\text{A}}$ above with $v=1$ and the (inverse) link $\eta(\cdot)$ suppressed (i.e., apply the test to the linear index $\mu_0(x)+\bw'\bgamma_0$ directly). 

Formally, the command \texttt{binstest} employs the test statistic
\[\dot{T}_p(x) = \frac{\widehat{\vartheta}_\bw^{(v)}(x) -  \frac{\partial^v}{\partial x^v}\eta(M_{{\bw}}(x;\widetilde{\btheta},\widetilde{\bgamma}))}{\sqrt{\widehat{\Omega}(x)/n}}, \qquad 0\leq v,s\leq p,\]
where $(\widetilde{\btheta}',\widetilde{\bgamma}')'$ are consistent estimates of $(\btheta',\bgamma_0')'$ under the null hypothesis (correct parametric specification), and are ``well behaved'' under the alternative hypothesis (parametric misspecification). The researcher needs to carefully choose a proper $v$ and decide if the test should be applied to $\eta(\mu_0(x)+\bw'\bgamma_0)$ or $\mu_0(x)+\bw'\bgamma_0$, following the discussion above. Then, a parametric specification hypothesis testing procedure is
\begin{equation}\label{eq:HT:specification-test}
\text{Reject } \dot{\mathsf{H}}_0 \qquad\text{ if and only if } \qquad\sup_{x} |\dot{T}_p(x)|\geq \cval,
\end{equation}
where $\cval=\inf\{c\in\mathbb{R}_+:\P[ \sup_{x} |\widehat{Z}_{p}(x)| \leq c \;| \;\bD ]\geq 1-\alpha \}$ is again computed by simulation from a standard normal random vector, conditional on the data $\bD$, as in the case of confidence bands already discussed. This testing procedure is an asymptotically valid $\alpha\%$-level test if the misspecification error is removed from the test statistic $\dot{T}_p(x)$.

The command \texttt{binstest} employs robust bias correction by default: first $p$ and $s$ are chosen, and the partitioning/binning scheme is selected by setting $J=J_\mathtt{IMSE}$ for these choices. Then, using this partitioning scheme, the testing procedure \eqref{eq:HT:specification-test} is implemented with the choice $p+q$ instead of $p$, with $q\geq1$. CCFF showed that, under regularity conditions, the resulting parametric specification testing approach controls Type I error with non-trivial power: for given $p$, $0\leq v,s\leq p$, and $J=J_\mathtt{IMSE}$,
\[\lim_{n\to\infty} \P\Big[ \sup_{x} \big|\dot{T}_{p+q}(x)\big| > \cval \Big] = \alpha, \qquad \text{under }\dot{\mathsf{H}}_0,\]
and
\[\lim_{n\to\infty} \P\Big[ \sup_{x} \big|\dot{T}_{p+q}(x)\big| > \cval \Big] = 1, \qquad \text{under }\dot{\mathsf{H}}_\text{A},\]
where $q\geq1$. This testing approach formalizes the intuitive idea that if the confidence band for $\vartheta_\bw^{(v)}(x)$ does not contain the hypothesized parametric fit entirely, then the parametric fit is incompatible with the data, i.e., the null should be rejected.

\subsubsection{Main implementation details}

The command \texttt{binstest} implements parametric specification testing in two ways. First, polynomial regression (parametric) specification testing is implemented directly via the option \texttt{testmodelpoly(P)}, where the null hypothesis is $m(x,\btheta)=\theta_0+x\theta_1+\cdots+x^{P}\theta_P$ and $\btheta=(\theta_0,\theta_1,\dots,\theta_P)'$ is estimated by least squares regression. For other parametrizations of $m(x,\btheta)$, the command takes as input an auxiliary array/database (\texttt{dta} in Stata, or data frame in \texttt{Python} and \texttt{R}) via the option \texttt{testmodelparfit({\it filename})} containing the following columns/variables: grid of evaluation points in one column, and fitted values $\eta^{(v)}(m(x,\widetilde{\btheta})+\bw'\widetilde{\bgamma})$ (over the evaluation grid) for each parametric model considered in other columns/variables. The ordering of these variables is arbitrary, but they have to follow a naming rule: the evaluation grid has the same name as the independent variable $x_i$, and the names of other variables storing fitted values take the form \texttt{binsreg\_fit*}. 

The binscatter (nonparametric) estimate used to construct the testing procedure is set by the options \texttt{testmodel(p s)} and \texttt{deriv(v)}, and the partitioning/binning scheme selected. If \texttt{testmodel(T)} or \texttt{testmodel()} is supplied, the default \texttt{testmodel(1 1)} is used unless the degree $p$ and smoothness $s$ selection is requested via the options \texttt{pselect()} and \texttt{sselect()} as described before. The option \texttt{nolink} can be used to specify if the test should be applied to the linear index directly.

\subsection{Nonparametric Shape Testing}

The second type of hypothesis tests implemented by the command \texttt{binstest} concern nonparametric testing of shape restrictions. For a choice of $v$, the null and alternative hypotheses of these testing  problems are:
\begin{align*}
\ddot{\mathsf{H}}_0:\quad \sup_{x}\vartheta_\bw^{(v)}(x) \leq 0, \qquad vs. \qquad 
\ddot{\mathsf{H}}_\text{A}:\quad \sup_{x} \vartheta_\bw^{(v)}(x) >0,
\end{align*}
that is, one-sided testing problem to the left. For example, negativity, monotonicity, and concavity of $\vartheta_\bw(x)$ correspond to $\vartheta_\bw(x)\leq 0$, $\vartheta_\bw^{(1)}(x)\leq0$, and $\vartheta_\bw^{(2)}(x)\leq0$, respectively. Of course, the analogous testing problem to the right is also implemented, but not discussed here to avoid unnecessary repetition.

The relevant Studentized test statistic for this class of testing problems is:
\[\ddot{T}_p(x) = \frac{\widehat{\vartheta}_\bw^{(v)}(x)}{\sqrt{\widehat{\Omega}(x)/n}}, \qquad 0\leq v,s\leq p.\]
Then, the testing procedure is:
\begin{equation}\label{eq:HT:specification-shape-right}
\text{Reject } \ddot{\mathsf{H}}_0 \qquad\text{ if and only if } \qquad\sup_{x} \ddot{T}_{p}(x) \geq \cval,
\end{equation}
with $\cval=\inf\{c\in\mathbb{R}_+:\P[ \sup_{x} \widehat{Z}_{p}(x) \leq c \;| \;\bD ]\geq 1-\alpha \}$. Misspecification errors of binscatter need to be taken into account in order to control Type I error. CCFF showed that for given $p$, $0\leq v,s\leq p$, and $J=J_\mathtt{IMSE}$ accordingly, then
\[\lim_{n\to\infty} \P\Big[ \sup_{x} \ddot{T}_{p+q}(x) > \cval \Big] \leq \alpha, \qquad \text{under }\ddot{\mathsf{H}}_0,\]
and
\[\lim_{n\to\infty} \P\Big[ \sup_{x} \ddot{T}_{p+q}(x) > \cval \Big] = 1, \qquad \text{under }\ddot{\mathsf{H}}_\text{A},\]
for any $q\geq1$, that is, using a robust bias correction approach. These results imply that the testing procedure \eqref{eq:HT:specification-shape-right} is an asymptotically valid hypothesis test provided that it is implemented with the choice $q\geq1$ after the IMSE-optimal partitioning/binning scheme for binscatter of order $p$ is selected.

\subsubsection{Main implementation details}

The command \texttt{binstest} implements one-sided and two-sided nonparametric shape restriction testing as follows. Option \texttt{testshapel(a)} implements one-sided testing to the left: $\ddot{\mathsf{H}}_0:\quad \sup_{x}\vartheta_\bw^{(v)}(x) \leq \mathtt{a}$. Option \texttt{testshaper(a)} for one-sided to the right: $\ddot{\mathsf{H}}_0:\quad \inf_{x}\vartheta_\bw^{(v)}(x) \geq \mathtt{a}$. Option \texttt{testshape2(a)} for two-sided testing: $\ddot{\mathsf{H}}_0:\quad \sup_{x} |\vartheta_\bw^{(v)}(x) - \mathtt{a}| = 0$. The constant \texttt{a} needs to be specified by the user. 

The binscatter (nonparametric) estimate used to construct the testing procedure is set by the options \texttt{testshape(p s)} and \texttt{deriv(v)}, and the chosen partitioning/binning scheme. If \texttt{testshape(T)} or \texttt{testshape()} is supplied, \texttt{testshape(1 1)} is used unless the degree $p$ and smoothness $s$ selection is requested via the options \texttt{pselect()} and \texttt{sselect()} as described before.

\subsection{Multi-Sample Estimation and Testing}

The package \textsf{Binsreg} also allows for comparisons of mean, quantile, and other regression functions across different groups (or treatment arms), which can be useful for estimation and inference of treatment effects that are heterogeneous in $x_i$, possibly after controlling for $\bw_i$. For each subsample defined by a group indicator variable, the parameter of interest can be defined as $\vartheta^{(v)}_{\bw,\ell}(x)$, which corresponds to the parameter in \eqref{eq:estimand-QMLE-PE} for specific subsample $\ell=0,1,2,\dots,L$.

For example, assuming that two sub-samples of the same size $n$ are available ($L=1$), one being a control group and the other a treatment group, all the methods discussed above can be applied to each subsample. Furthermore, the null hypothesis of no heterogeneous treatment effect is: $\mathsf{H}^\Delta_0: \vartheta^{(v)}_{\bw,0}(x)=\vartheta^{(v)}_{\bw,1}(x)$ for all $x\in\mathcal{X}$, which captures the idea of no (heterogeneous in $x_i$) treatment effect across the two groups. A natural test statistic is:
\[T^\Delta_p(x) = \frac{\widehat{\vartheta}^{(v)}_{{\bw},1}(x) - \widehat{\vartheta}^{(v)}_{{\bw},0}(x)}{\sqrt{\widehat{\Omega}_1(x)/n+\widehat{\Omega}_0(x)/n}}, \qquad 0\leq v,s\leq p,\]
which compares the pairwise difference between the two groups, where $\widehat{\Omega}_{\ell}(x)$ is the variance estimator (of $\widehat{\vartheta}^{(v)}_{{\bw},\ell}(x)$) for the subsample $\ell=0,1$. The testing procedure is:
\begin{equation}\label{eq:HT:specification-pairwise}
	\text{Reject } \mathsf{H}^\Delta_0 \qquad\text{ if and only if } \qquad\sup_{x} \left| T_p^\Delta(x) \right| \geq \cval,
\end{equation}
with the critical value obtained as before via Gaussian approximations (resampling from a normal random vector conditional on the data). As discussed before, in practice the robust bias-corrected test statistics $T_{p+q}^\Delta(x)$ is used to eliminate misspecification bias and obtain a valid hypothesis testing procedure.

All the ideas and results above also apply to pairwise comparisons across multi-samples. In particular, estimation, uncertainty quantification and hypothesis testing can be conducted for each subsample at the time, and then hypothesis testing for pairwise comparisons can also be implemented following the results above. CCFF provided all the necessary theoretical background. Importantly, concerns regarding the choice of evaluation point $\bw$ also apply to the multi-sample testing problems: researchers need to be careful when implementing the tests and interpreting the results.

\subsubsection{Main implementation details}

Estimation and uncertainty quantification across subsamples is done using the estimation commands (\texttt{binsreg}, \texttt{binslogit}, \texttt{binsprobit}, and \texttt{binsqreg}) via the option \texttt{by()}. In addition, the command \texttt{binspwc} implements formal hypothesis testing for pairwise comparisons for the null hypothesis $\mathsf{H}^\Delta_0$ (and analogous one-sided problems).

\subsection{Extensions and Other Implementation Details}\label{section:Extensions and Other Implementation Details}

The package \textsf{Binsreg} is implemented using the base commands in the statistical software. For example, in \texttt{Stata}, \texttt{binsreg} relies on \texttt{regress} (or \texttt{reghdfe} if that option is selected), \texttt{binslogit} relies on \texttt{logit}, \texttt{binsprobit} relies on \texttt{probit}, and \texttt{binsqreg} relies on \texttt{qreg} (or \texttt{bsqreg} if bootstrapping-based standard error is selected). Furthermore, the testing commands (\texttt{binstest} and \texttt{binspwc}) also employ base commands whenever possible. This approach may sacrifice some speed of implementation, but improves substantially in terms of stability and replicability. Importantly, essentially most options available in the base commands are available in the package \textsf{Binsreg}.

This section reviews some specific extensions and other numerical issues of the package \textsf{Binsreg} and discusses related choices made for implementation, all of which can affect speed and/or robustness of the package. 

\subsubsection{Other metrics}

All the results presented above employ the uniform norm, that is, focus on the the largest deviation on the support of a function. See, for example, $\dot{\mathsf{H}}_0$, $\ddot{\mathsf{H}}_0$, and $\mathsf{H}^\Delta_0$. Our results also apply to other metrics, such as the $L_\mathfrak{p}$ metric. In such a case, the null hypotheses, the corresponding statistics and simulated critical values will focus on an integral computation of the function of interest. For example, $\dot{\mathsf{H}}_0$ is replaced by
\[\int \Big|\vartheta_\bw^{(v)}(x) - \frac{\partial^v}{\partial x^v}\eta(M_\bw(x;\btheta,\bgamma_0))\Big|^\mathfrak{p} dx = 0, \quad \text{ for some } \btheta,\]
where $\mathfrak{p}$ is some positive integer no less than 1 (typically $\mathfrak{p}=2$ for squared deviations), and the corresponding critical value simulation takes the form $\cval=\inf\{c\in\mathbb{R}_+:\P[ \int  |\widehat{Z}_{p}(x)|^\mathfrak{p} dx \leq c \;| \;\bD ]\geq 1-\alpha \}$. Analogous modifications are done for other hypothesis tests. Note that by construction, the $L_\mathfrak{p}$ metric measures the integrated absolute deviation, and thus should be applied to two-sided tests only. The choice of metric is implemented in each testing command via the option \texttt{lp()}.

\subsubsection{Mass points and minimum effective sample size}

The package \textsf{Binsreg} incorporates specific implementation decisions to deal with mass points in the distribution of the independent variable $x_i$. The number of distinct values of $x_i$, denoted by $N$, is taken as the effective sample size as opposed to the total number of observations $n$. If $x_i$ is continuously distributed, then $N=n$. However, in many applications, $N$ can be substantially smaller than $n$, and this affects some of the implementations in the package.

First, assume that $J$ is set by the user (via the option \texttt{nbins(J)}). Then, given the choice $J$, the commands \texttt{binsreg}, \texttt{binslogit}, \texttt{binsprobit}, \texttt{binsqreg}, \texttt{binstest}, and \texttt{binspwc} perform a degrees of freedom check to decide whether the $x_i$ data exhibit enough variation. Specifically, given $p$ and $s$ set by the option \texttt{dots(p s)} or \texttt{bins(p s)}, these  commands check whether $N > N_2 + (p+1)J-(J-1)s$ with $N_2=30$ by default. If this check is not passed, then the package \textsf{Binsreg} regards the data as having ``too little'' variation in $x_i$, and turns off all nonparametric estimation and inference results based on large sample approximations. Thus, in this extreme case, the command  \texttt{binsreg} (or \texttt{binslogit}, \texttt{binsprobit}, \texttt{binsqreg}) only allows for \texttt{dots(0 0)}, \texttt{ci(0 0)}, and \texttt{polyreg(P)} for any $P+1<N$, while the command \texttt{binstest} (or \texttt{binspwc}) does not return any results and issues a warning message instead.

If, on the other hand, for given $J$, the numerical check $N > N_2 + (p+1)J-(J-1)s$ is passed, then all nonparametric methods implemented by the commands \texttt{binsreg}, \texttt{binslogit}, \texttt{binsprobit}, \texttt{binsqreg}, \texttt{binstest}, and  \texttt{binspwc} become available. However, before implementing each method (\texttt{dots(p s)}, \texttt{lines(p s)}, \texttt{ci(p s)}, \texttt{cb(p s)}, \texttt{polyreg(P)}, and the hypothesis testing procedures), a degrees of freedom check is performed in each individual case. Specifically, each nonparametric procedure is implemented only if $N > N_2 + (p+1)J-(J-1)s$, where recall that $p$ and $s$ may change from one procedure to the next. 

Second, as discussed above, whenever $J$ is not set by the user via the option \texttt{nbins()}, the command \texttt{binsregselect} is employed to select $J$ in a data-driven way, provided there is enough variation in $x_i$. To determine the latter, an initial degrees of freedom check is performed to assess whether $J$ selection is possible or, alternatively, if the unique values of $x_i$ should be used as bins directly. Specifically, if $N > N_1 + p + 1$, with $p$ set by the option \texttt{dots(p s)} (or \texttt{bins(p s)}) and $N_1=20$ by default, then the data are deemed appropriate for ROT selection of $J$ via the command \texttt{binsregselect}, and hence $\widehat{J}_\mathtt{ROT}$ is implemented. If, in addition, $N > N_2 + (p+1)\widehat{J}_\mathtt{ROT}-(\widehat{J}_\mathtt{ROT}-1)s$, then $\widehat{J}_\mathtt{DPI}$ is also implemented whenever requested. Furthermore, the command \texttt{binsregselect} employs the following alternative formula for $J$ selection:
\[J_\texttt{IMSE} = \left\lceil \left(\frac{2(p-v+1)\mathscr{B}_n(p,s,v)}{(1+2v)\mathscr{V}_n(p,s,v)}\right)^{\frac{1}{2p+3}} \; N^{\frac{1}{2p+3}} \right\rceil,\]
with a slightly different constant $\mathscr{V}_n(p,s,v)$, taking into account the frequency of data at each mass point. All other estimators in the package \textsf{Binsreg}, including bias and standard error estimators, automatically adapt to the presence of mass points. Once the final $J$ is estimated, the degrees of freedom checks discussed in the previous paragraphs are performed based on this choice.

If $J$ is not set by the user and $N \leq N_1 + p + 1$, so that not even ROT estimation of $J$ is possible, then $N$ is taken as ``too small.'' In this extreme case, the package \textsf{Binsreg} sets $J=N$ and constructs a partitioning/binning structure with each bin containing one unique value of $x_i$. In other words, the support of the raw data is taken as the binning structure itself. In this extreme case, the follow-up degrees of freedom checks based on the formula $N > N_2 + (p+1)J-(J-1)s$ fail by construction, and hence the nonparametric methods are turned off as explained above.

Finally, the specific numerical checks and corresponding adjustments mentioned in this subsection can be modified or omitted. This is controlled by two main options: \texttt{dfcheck()} and \texttt{masspoints()}, respectively. First, the default cutoff points $N_1$ and $N_2$, corresponding to the degrees of freedom checks for parametric global polynomial regression and nonparametric binscatter, respectively, can be modified using the option \texttt{dfcheck($\mathtt{N_1}$ $\mathtt{N_2}$)}. Second, the option \texttt{masspoints()} controls how the package \textsf{Binsreg} handles the presence of mass points (i.e., repeated values) in $x_i$. Specifically, setting \texttt{masspoints(noadjust)} omits mass point checks and the corresponding effective sample size adjustments, that is, it sets $N=n$ and ignores the presence of mass points in $x_i$ (if any). Setting \texttt{masspoints(nolocalcheck)} omits within-bin mass point checks, but still performs global mass point checks and adjustments. The option \texttt{masspoints(off)} corresponds to setting both \texttt{masspoints(noadjust)} and \texttt{masspoints(nolocalcheck)} simultaneously. Finally, setting \texttt{masspoints(veryfew)} forces the package to proceed as if $N$ is so small that all checks are failed, thereby treating $x_i$ as if it has very few distinct values.
	
\subsubsection{Clustered data and minimum effective sample size}

As discussed in CCFF, the main methodological results for binscatter can be extended to accommodate clustered data. All three commands in the package \textsf{Binsreg} allow for clustered data via the option \texttt{vce()}. In this case, the number of clusters $G$ is taken as the effective sample size, assuming $N=n$ (see below for the other case). The only substantive change occurs in the command \texttt{binsregselect}, which now employs the following alternative formula for $J$ selection:
\[J_\texttt{IMSE} = \left\lceil \left(\frac{2(p-v+1)\mathscr{B}_n(p,s,v)}{(1+2v)\mathscr{V}_n(p,s,v)}\right)^{\frac{1}{2p+3}} \; G^{\frac{1}{2p+3}} \right\rceil,\]
with a variance constant $\mathscr{V}_n(p,s,v)$ accounting for the clustered structure of the data. Accordingly, cluster-robust variance estimators are used in this case.

\subsubsection{Minimum effective sample size}

The package \textsf{Binsreg} requires some minimal variation in $x_i$ in order to successfully implement nonparametric methods based on large sample approximations. The minimal variation is captured by the number of distinct values on the support of $x_i$, denoted by $N$, and the number of clusters, denoted by $G$. Thus, all three commands in the package perform degrees of freedom numerical checks using $\min\{n,N,G\}$ as the general definition of effective sample size, and proceeding as explained above for the case of mass points in the distribution of $x_i$.

\subsection{Increasing Speed of Execution}\label{section:Increasing Speed of Execution}

The package \textsf{Binsreg} offers a large array of options and methods, many of which involve nonlinear estimation and/or simulations, thereby slowing down its speed of execution. Furthermore, in order to improve the stability and replicability of the package, it implements several robustness checks that may further decrease execution speed, particularly in settings with ultra-large datasets. There are, however, several options and approaches that could be used to improve the speed of execution of the package \textsf{Binsreg}.

\begin{enumerate}
	\item \textit{Sorted data}. The core implementations of the package \textsf{Binsreg} employ several algorithms and procedures that require sorted data along the $x$ dimension. If the provided data is not sorted, then the package begins by sorting the data, which slows down the execution (particularly in large datasets).
	\begin{itemize}
		\item Speed improvement: provide sorted data in $x$, which may substantially increase execution speed (particularly in ultra-large datasets).
	\end{itemize}
	
	\item \textit{Data Distribution}. The methods implemented in the package \textsf{Binsreg} were developed for continuously distributed data with ``enough" variation (e.g., enough degrees of freedom within and across bins, appropriate rank conditions for Gram matrices, etc.). Because empirical work may involve data with mass points and/or other irregularities that can make the default methods fail, the package \textsf{Binsreg} implements a series of robustness checks before execution (see above for details). 
	\begin{itemize}
		\item Speed improvement: use option \texttt{masspoints(off)} whenever $x$ is known to be (close to) continuously distributed and the data exhibits ``enough" regularity. 
	\end{itemize}
	
	\item \textit{Number of Bins Selection}. The package \textsf{Binsreg} selects the number of bins $J$ in a multi-step, data-driven and optimal way, whenever the user does not provide a selection manually (via the options \texttt{nbins()} or \texttt{bynbins()}). In large datasets, estimating $J$ may be time consuming. 
	\begin{itemize}
		\item Speed improvement: provide $J$ manually or use option \texttt{randcut({\it numeric})} to speed up the process.
	\end{itemize}
	
	\item \textit{Gtools}. The package \textsf{Binsreg} is open source and, by default, relies exclusively on base commands and functions in \texttt{Stata} (as well as in \texttt{Python} and \texttt{R}). However, some parts of this algorithm (e.g., \texttt{pctile}) may be slow in large datasets.
	\begin{itemize}
		\item Speed improvement: employ the community-distributed package \textsf{gtools} \citep{Caceres_2024_gtools} via the option \texttt{usegtools(on)}. This community-distributed package needs to be installed separately by the user.
	\end{itemize}
	
	\item \textit{Other Possibilities}. The package \textsf{Binsreg} offers several other options for increasing speed of execution. First, the community-distributed package \textsf{reghdfe} \citep{Correia-Constantine_2024_reghdfe} could be used when employing the \texttt{binsreg} command. Second, for uncertainty quantification and inference, the number of simulations (option \texttt{nsims()}) and the number of grid points for simulation (option \texttt{simsgrid}()) can be decreased to improve speed, which could offer a good alternative for preliminary exploration. Finally, for ultra-large datasets, it may be advisable to begin exploratory analysis with a random sample of the data, if the goal is to increase speed of execution of the package \textsf{Binsreg}.
	
\end{enumerate}

\section{Illustration of Methods}\label{section:illustration}

We illustrate the package \textsf{Binsreg} using a simulated dataset, which is available in the file \texttt{binscatter\_simdata.dta}. In this dataset, \texttt{y} is the outcome variable, \texttt{x} is the independent variable for binning, \texttt{w} is a continuously distributed covariate, and \texttt{t} is a binary covariate, and \texttt{id} is a group identifier. Summary statistics of the simulated data are as follows.

{\fontsize{8}{8}\selectfont\begin{stlog}[auto]. use binsreg_simdata, clear
{\smallskip}
. sum
{\smallskip}
    Variable {\VBAR}        Obs        Mean    Std. dev.       Min        Max
\HLI{13}{\PLUS}\HLI{57}
           x {\VBAR}      1,000    .4907072    .2932553   .0002281   .9985808
           w {\VBAR}      1,000    .0120224    .5799381  -.9993055   .9973198
           t {\VBAR}      1,000        .515     .500025          0          1
          id {\VBAR}      1,000       250.5    144.4095          1        500
           y {\VBAR}      1,000    .5283884    1.727878  -5.159858   5.751276
\HLI{13}{\PLUS}\HLI{57}
           d {\VBAR}      1,000         .45    .4977427          0          1
{\smallskip}
\end{stlog}}

\subsection{Estimation, Uncertainty Quantification and Plotting}

The basic syntax for \texttt{binsreg} is the following:

{\fontsize{8}{8}\selectfont\begin{stlog}[auto]. binsreg y x w
Sorting dataset on x...
Note: This step is omitted if dataset already sorted by x.
{\smallskip}
Binscatter plot
Bin selection method: IMSE-optimal plug-in choice (select \# of bins)
Placement: Quantile-spaced
Derivative: 0
{\smallskip}
\HLI{30}{\TOPT}\HLI{15}
\# of observations             {\VBAR}    1000
\# of distinct values          {\VBAR}    1000
\# of clusters                 {\VBAR}       .
\HLI{30}{\PLUS}\HLI{15}
Bin/Degree selection:         {\VBAR} 
         Degree of polynomial {\VBAR}       0
  \# of smoothness constraints {\VBAR}       0
                    \# of bins {\VBAR}      21
                 imse, bias{\caret}2 {\VBAR}   5.420
                   imse, var. {\VBAR}   1.192
\HLI{30}{\BOTT}\HLI{15}
{\smallskip}
\HLI{9}{\TOPT}\HLI{30}
         {\VBAR}      p       s       df
\HLI{9}{\PLUS}\HLI{30}
 dots    {\VBAR}      0       0       21
\HLI{9}{\BOTT}\HLI{30}
\end{stlog}}

\begin{figure}[h]
	\centering
	\includegraphics[scale=1]{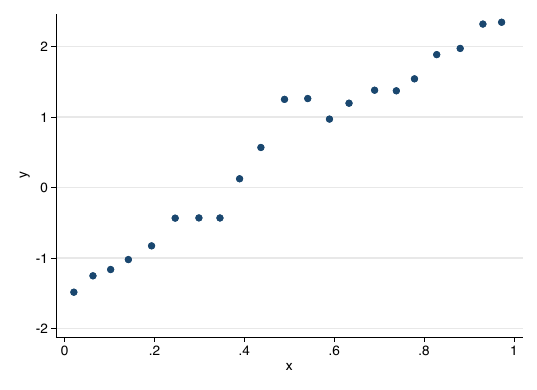}
	\caption{Canonical Binned Scatter Plot.}\label{fig:dots}
\end{figure}

The main output is a binned scatter plot as shown in Figure \ref{fig:dots}. By default, the (nonparametric) mean relationship between \texttt{y} and \texttt{x} is approximated by piecewise constants (\texttt{dots(0 0)}). Each dot in the figure represents the point estimate corresponding to each bin, which is the canonical binscatter plot. The number of bins, whenever not specified, is automatically selected via the companion command \texttt{binsregselect}. In this case, $21$ bins are used. Other useful information is also reported, including total sample size, the number of distinct values of \texttt{x}, bin selection results, and the degrees of freedom of the statistical model(s) employed. 

By default, the command \texttt{binsreg} evaluates and plots the regression function of interest $\vartheta^{(v)}_\bw(x)$ at the mean of the additional covariates $\bw_i$, i.e., $\bw=\bar{\bw}$. Users may specify a different value of $\bw$, for example, the empirical median of each component in $\bw_i$, via the option \texttt{at()}:

\texttt{. binsreg y x w, at(median)}

Users may also save the values of the additional covariates at which the binscatter estimate is evaluated in another file, and then specify the file name in the option \texttt{at()}. For example,

{\fontsize{8}{8}\selectfont\begin{stlog}[auto]. tempfile evalcovar
{\smallskip}
. preserve
{\smallskip}
. clear
{\smallskip}
. set obs 1
Number of observations ({\bftt{_N}}) was 0, now 1.
{\smallskip}
. gen w=0.2
{\smallskip}
. gen t=1
{\smallskip}
. save `evalcovar', replace
(file{\bftt{ /var/folders/0b/h0wl9g7d3s3dr9d_vm0qts9r0000gn/T//S_37285.000001}} not found)
file{\bftt{ /var/folders/0b/h0wl9g7d3s3dr9d_vm0qts9r0000gn/T//S_37285.000001}} saved as .dta
    format
{\smallskip}
. restore
{\smallskip}
. binsreg y x w i.t, at(`evalcovar')
Sorting dataset on x...
Note: This step is omitted if dataset already sorted by x.
{\smallskip}
Binscatter plot
Bin selection method: IMSE-optimal plug-in choice (select \# of bins)
Placement: Quantile-spaced
Derivative: 0
{\smallskip}
\HLI{30}{\TOPT}\HLI{15}
\# of observations             {\VBAR}    1000
\# of distinct values          {\VBAR}    1000
\# of clusters                 {\VBAR}       .
\HLI{30}{\PLUS}\HLI{15}
Bin/Degree selection:         {\VBAR} 
         Degree of polynomial {\VBAR}       0
  \# of smoothness constraints {\VBAR}       0
                    \# of bins {\VBAR}      22
                 imse, bias{\caret}2 {\VBAR}   4.736
                   imse, var. {\VBAR}   0.974
\HLI{30}{\BOTT}\HLI{15}
{\smallskip}
\HLI{9}{\TOPT}\HLI{30}
         {\VBAR}      p       s       df
\HLI{9}{\PLUS}\HLI{30}
 dots    {\VBAR}      0       0       22
\HLI{9}{\BOTT}\HLI{30}
\end{stlog}}

In this case, we control for a continuous variable \texttt{w} and a dummy variable generated based on the binary covariate \texttt{t}. We evaluate the binscatter estimate at \texttt{w=0.2} and \texttt{t=1}, and these values are saved in the temporary file \texttt{`evalcovar'} in advance.

Users may specify the number of bins manually rather than relying on the automatic data-driven procedures. For example, a popular ad-hoc choice in practice is setting $J=20$ quantile-spaced bins:

{\fontsize{8}{8}\selectfont\begin{stlog}[auto]. binsreg y x w, nbins(20) polyreg(1)
Sorting dataset on x...
Note: This step is omitted if dataset already sorted by x.
Note: When additional covariates w are included, the polynomial fit may not always be clos
> e to the binscatter fit.
{\smallskip}
Binscatter plot
Bin selection method: User-specified
Placement: Quantile-spaced
Derivative: 0
{\smallskip}
\HLI{30}{\TOPT}\HLI{15}
\# of observations             {\VBAR}    1000
\# of distinct values          {\VBAR}    1000
\# of clusters                 {\VBAR}       .
\HLI{30}{\PLUS}\HLI{15}
Bin/Degree selection:         {\VBAR} 
         Degree of polynomial {\VBAR}       0
  \# of smoothness constraints {\VBAR}       0
                    \# of bins {\VBAR}      20
\HLI{30}{\BOTT}\HLI{15}
{\smallskip}
\HLI{9}{\TOPT}\HLI{30}
         {\VBAR}      p       s       df
\HLI{9}{\PLUS}\HLI{30}
 dots    {\VBAR}      0       0       20
 polyreg {\VBAR}      1       NA      2
\HLI{9}{\BOTT}\HLI{30}
\end{stlog}}

The option \texttt{polyreg(1)} adds a linear prediction line to the canonical binscatter plot, but the resulting binned scatter plot is not reported here to conserve space.

\begin{figure}[h]
	\begin{center}\caption{Binned Scatter Plot with Lines, Confidence Intervals and Bands.\label{fig:binscatter_all}}
		\vspace{-.1in}
		\begin{subfigure}{0.495\textwidth}
			\includegraphics[width=\textwidth]{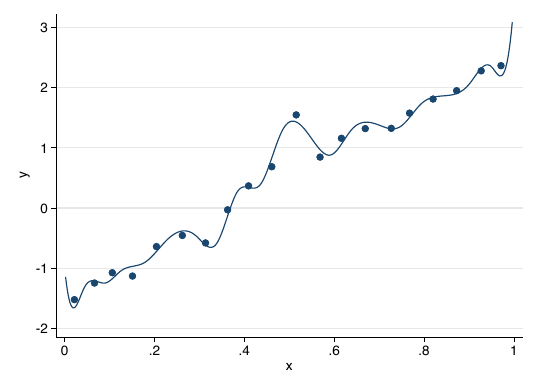}
			\caption{Add cubic $B$-spline fit}
		\end{subfigure}
		\begin{subfigure}{0.495\textwidth}
			\includegraphics[width=\textwidth]{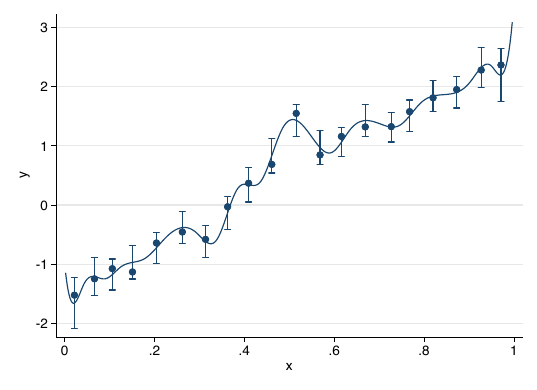}
			\caption{Add confidence intervals}
		\end{subfigure}\\
		\begin{subfigure}{0.495\textwidth}
			\includegraphics[width=\textwidth]{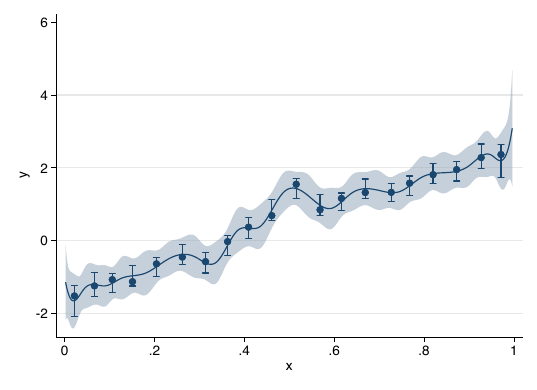}
			\caption{Add confidence band}
		\end{subfigure}
		\begin{subfigure}{0.495\textwidth}
			\includegraphics[width=\textwidth]{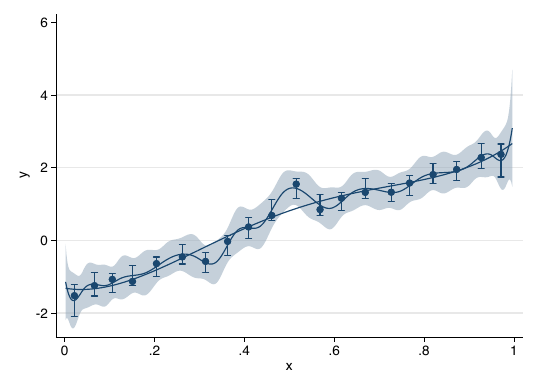}
			\caption{Add a polynomial fit of degree $4$}
		\end{subfigure}
	\end{center}
\end{figure}

The command \texttt{binsreg} allows users to add a binscatter-based line approximating the unknown regression function, pointwise confidence intervals, a uniform confidence band, and a global polynomial regression approximation. For example, the following syntax cumulatively adds in four distinct plots a fitted line, confidence intervals and a confidence band, all three based on cubic $B$-splines, and also a fitted line based on a global polynomial of degree $4$. The results are shown in Figure \ref{fig:binscatter_all}.

{\fontsize{8}{8}\selectfont\begin{stlog}[auto]. qui binsreg y x w, nbins(20) dots(0,0) line(3,3)
{\smallskip}
. qui binsreg y x w, nbins(20) dots(0,0) line(3,3) ci(3,3)
{\smallskip}
. qui binsreg y x w, nbins(20) dots(0,0) line(3,3) ci(3,3) cb(3,3)
{\smallskip}
. qui binsreg y x w, nbins(20) dots(0,0) line(3,3) ci(3,3) cb(3,3) polyreg(4)
{\smallskip}
\end{stlog}}

By construction, a cubic $B$-spline fit is a piecewise cubic polynomial function which is continuous, and has continuous first- and second-order derivatives. Thus, the prediction line and confidence band generated are quite smooth. In this case, it is arguably under-smoothed because of the ``large'' choice of $J=20$. The degree and smoothness of polynomials can be changed by adjusting the values of \texttt{p} and \texttt{s} in the options \texttt{dots()}, \texttt{line()}, \texttt{ci()}, and \texttt{cb()}.

The command \texttt{binsreg} also allows for the standard \texttt{vce} options, factor variables, and \texttt{twoway} graph options, among other features. This is illustrated in the following code:

{\fontsize{8}{8}\selectfont\begin{stlog}[auto]. binsreg y x w i.t, dots(0,0) line(3,3) ci(3,3) cb(3,3) polyreg(4) ///
>                    vce(cluster id) savedata(output/graphdat) replace ///
>                                    title("Binned Scatter Plot") 
Sorting dataset on x...
Note: This step is omitted if dataset already sorted by x.
Note: When additional covariates w are included, the polynomial fit may not always be clos
> e to the binscatter fit.
Note: Setting at least nsims(2000) and simsgrid(50) is recommended to obtain the final res
> ults.
{\smallskip}
Binscatter plot
Bin selection method: IMSE-optimal plug-in choice (select \# of bins)
Placement: Quantile-spaced
Derivative: 0
Output file: output/graphdat.dta
{\smallskip}
\HLI{30}{\TOPT}\HLI{15}
\# of observations             {\VBAR}    1000
\# of distinct values          {\VBAR}    1000
\# of clusters                 {\VBAR}     500
\HLI{30}{\PLUS}\HLI{15}
Bin/Degree selection:         {\VBAR} 
         Degree of polynomial {\VBAR}       0
  \# of smoothness constraints {\VBAR}       0
                    \# of bins {\VBAR}      20
                 imse, bias{\caret}2 {\VBAR}   3.588
                   imse, var. {\VBAR}   0.494
\HLI{30}{\BOTT}\HLI{15}
{\smallskip}
\HLI{9}{\TOPT}\HLI{30}
         {\VBAR}      p       s       df
\HLI{9}{\PLUS}\HLI{30}
 dots    {\VBAR}      0       0       20
 line    {\VBAR}      3       3       23
 CI      {\VBAR}      3       3       23
 CB      {\VBAR}      3       3       23
 polyreg {\VBAR}      4       NA      5
\HLI{9}{\BOTT}\HLI{30}
\end{stlog}}

Specifically, a dummy variable based on the binary covariate \texttt{t} is added to the estimation, standard errors are clustered at the group level indicator \texttt{id}, and a graph title is added to the resulting binned scatter plot. Note that any unrecognized options for the command \texttt{binsreg} will be understood as \texttt{twoway} options and therefore appended to the final plot command. Thus, users may easily modify, for example, axis properties, legends, etc. The option \texttt{savedata(graphdat)} saves the underlying data used in the binned scatter plot in the file \texttt{graphdat.dta}.

\begin{figure}[h]
	\centering 
	\includegraphics[scale=1]{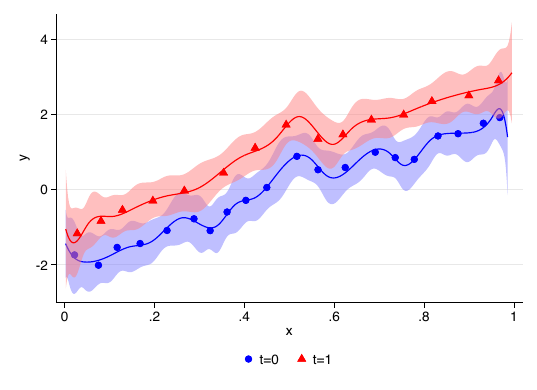}
	\caption{Binned Scatter Plot: Group Comparison}\label{fig:2groups}
\end{figure}

In addition, the command \texttt{binsreg} can be used for subgroup analysis. The following command implements binscatter estimation and inference across two subgroups separately, defined by the variable \texttt{t}, and then produces a common binned scatter plot (Figure \ref{fig:2groups}):

{\fontsize{8}{8}\selectfont\begin{stlog}[auto]. binsreg y x w, by(t) dots(0,0) line(3,3) cb(3,3) ///
>                bycolors(blue red) bysymbols(O T) 
Sorting dataset on x...
Note: This step is omitted if dataset already sorted by x.
Note: Setting at least nsims(2000) and simsgrid(50) is recommended to obtain the final res
> ults.
{\smallskip}
Binscatter plot
Bin selection method: IMSE-optimal plug-in choice (select \# of bins)
Placement: Quantile-spaced
Derivative: 0
{\smallskip}
Group: t = 0
\HLI{30}{\TOPT}\HLI{15}
\# of observations             {\VBAR}     485
\# of distinct values          {\VBAR}     485
\# of clusters                 {\VBAR}       .
\HLI{30}{\PLUS}\HLI{15}
Bin/Degree selection:         {\VBAR} 
         Degree of polynomial {\VBAR}       0
  \# of smoothness constraints {\VBAR}       0
                    \# of bins {\VBAR}      20
                 imse, bias{\caret}2 {\VBAR}   7.092
                   imse, var. {\VBAR}   0.941
\HLI{30}{\BOTT}\HLI{15}
{\smallskip}
\HLI{9}{\TOPT}\HLI{30}
         {\VBAR}      p       s       df
\HLI{9}{\PLUS}\HLI{30}
 dots    {\VBAR}      0       0       20
 line    {\VBAR}      3       3       23
 CB      {\VBAR}      3       3       23
\HLI{9}{\BOTT}\HLI{30}
Note: Setting at least nsims(2000) and simsgrid(50) is recommended to obtain the final res
> ults.
{\smallskip}
{\smallskip}
Group: t = 1
\HLI{30}{\TOPT}\HLI{15}
\# of observations             {\VBAR}     515
\# of distinct values          {\VBAR}     515
\# of clusters                 {\VBAR}       .
\HLI{30}{\PLUS}\HLI{15}
Bin/Degree selection:         {\VBAR} 
         Degree of polynomial {\VBAR}       0
  \# of smoothness constraints {\VBAR}       0
                    \# of bins {\VBAR}      15
                 imse, bias{\caret}2 {\VBAR}   2.861
                   imse, var. {\VBAR}   0.955
\HLI{30}{\BOTT}\HLI{15}
{\smallskip}
\HLI{9}{\TOPT}\HLI{30}
         {\VBAR}      p       s       df
\HLI{9}{\PLUS}\HLI{30}
 dots    {\VBAR}      0       0       15
 line    {\VBAR}      3       3       18
 CB      {\VBAR}      3       3       18
\HLI{9}{\BOTT}\HLI{30}
\end{stlog}}

Figure \ref{fig:2groups} highlights a difference across the two subgroups defined by the variable \texttt{t}, which corresponds to the fact that our simulated data add a $1$ to the outcome variable for those units with $\mathtt{t=1}$. The colors, symbols, and line patterns in Figure \ref{fig:2groups} can be modified via the options \texttt{bycolors()}, \texttt{bysymbols()}, and \texttt{bylpatterns()}. When the number of bins is unspecified, the command \texttt{binsreg} selects the number of bins for each subsample separately, via the companion command \texttt{binsregselect}. This means that, by default, the choice of binning/partitioning structure will be different across subgroups in general. However, if the option \texttt{samebinsby} is specified, then a common binning scheme for all subgroups is constructed based on the full sample.

As described before, sometimes one would like to keep the number of bins $J$ fixed and select the degree/smoothness of the polynomial instead. The following snippet illustrates how to implement this procedure:

{\fontsize{8}{8}\selectfont\begin{stlog}[auto]. binsreg y x w, nbins(20) line(T) ci(T) cb(T) pselect(0/3)
Sorting dataset on x...
Note: This step is omitted if dataset already sorted by x.
Note: Setting at least nsims(2000) and simsgrid(50) is recommended to obtain the final res
> ults.
{\smallskip}
Binscatter plot
Bin selection method: IMSE-optimal plug-in choice (select degree and smoothness)
Placement: Quantile-spaced
Derivative: 0
{\smallskip}
\HLI{30}{\TOPT}\HLI{15}
\# of observations             {\VBAR}    1000
\# of distinct values          {\VBAR}    1000
\# of clusters                 {\VBAR}       .
\HLI{30}{\PLUS}\HLI{15}
Bin/Degree selection:         {\VBAR} 
         Degree of polynomial {\VBAR}       0
  \# of smoothness constraints {\VBAR}       0
                    \# of bins {\VBAR}      20
                 imse, bias{\caret}2 {\VBAR}   5.422
                   imse, var. {\VBAR}   1.181
\HLI{30}{\BOTT}\HLI{15}
{\smallskip}
\HLI{9}{\TOPT}\HLI{30}
         {\VBAR}      p       s       df
\HLI{9}{\PLUS}\HLI{30}
 dots    {\VBAR}      0       0       20
 line    {\VBAR}      0       0       20
 CI      {\VBAR}      1       1       21
 CB      {\VBAR}      1       1       21
\HLI{9}{\BOTT}\HLI{30}
\end{stlog}}

Here we let $J=20$ and select the degree of polynomial $p$ within the specified range $\{0, 1, 2, 3\}$. The resulting optimal $p$ is 0. Accordingly, we construct ``dots'' and ``line'' based on piecewise constant estimates, and  confidence intervals and a confidence band based on linear spline estimates.

The accompanying replication files include other illustrations. For example:
\begin{itemize}
	\item Inference based on asymptotic variance formula:\\
	      \texttt{. binsreg y x w i.t, dots(0 0) line(3 3) ci(3 3) cb(3 3) polyreg(4) vce(cluster id) asyvar(on)}
	\item Using the community-contributed module \texttt{reghdfe}:\\
          \texttt{. binsreg y x w, absorb(t) dots(0 0) line(3 3) ci(3 3) cb(3 3) polyreg(4)}
	\item Turning off data distribution robustness checks and using the community-contributed module \texttt{gtools}:\\
          \texttt{. binsreg y x w, masspoints(off) usegtools(on)}
\end{itemize}

Next, we illustrate the command \texttt{binsqreg} for estimation and uncertainty quantification using quantile regression binscatter methods. The following code looks at the conditional $25$-th quantile of the outcome variable.

{\fontsize{8}{8}\selectfont\begin{stlog}[auto]. binsqreg y x w, quantile(0.25)
Sorting dataset on x...
Note: This step is omitted if dataset already sorted by x.
{\smallskip}
Binscatter plot, quantile
Bin selection method: IMSE-optimal plug-in choice (select \# of bins)
Placement: Quantile-spaced
Derivative: 0
{\smallskip}
\HLI{30}{\TOPT}\HLI{15}
\# of observations             {\VBAR}    1000
\# of distinct values          {\VBAR}    1000
\# of clusters                 {\VBAR}       .
\HLI{30}{\PLUS}\HLI{15}
Bin/Degree selection:         {\VBAR} 
         Degree of polynomial {\VBAR}       0
  \# of smoothness constraints {\VBAR}       0
                    \# of bins {\VBAR}      21
                 imse, bias{\caret}2 {\VBAR}   5.420
                   imse, var. {\VBAR}   1.192
\HLI{30}{\BOTT}\HLI{15}
{\smallskip}
\HLI{9}{\TOPT}\HLI{30}
         {\VBAR}      p       s       df
\HLI{9}{\PLUS}\HLI{30}
 dots    {\VBAR}      0       0       21
\HLI{9}{\BOTT}\HLI{30}
\end{stlog}}

By default, quantile regression methods employ an analytic variance estimator formula, which may not perform well in applications. A more robust alternative is employing bootstrap methods:

{\fontsize{8}{8}\selectfont\begin{stlog}[auto]. binsqreg y x w, quantile(0.25) ci(3 3) vce(bootstrap, reps(100))
Sorting dataset on x...
Note: This step is omitted if dataset already sorted by x.
{\smallskip}
Binscatter plot, quantile
Bin selection method: IMSE-optimal plug-in choice (select \# of bins)
Placement: Quantile-spaced
Derivative: 0
{\smallskip}
\HLI{30}{\TOPT}\HLI{15}
\# of observations             {\VBAR}    1000
\# of distinct values          {\VBAR}    1000
\# of clusters                 {\VBAR}       .
\HLI{30}{\PLUS}\HLI{15}
Bin/Degree selection:         {\VBAR} 
         Degree of polynomial {\VBAR}       0
  \# of smoothness constraints {\VBAR}       0
                    \# of bins {\VBAR}      21
                 imse, bias{\caret}2 {\VBAR}   5.420
                   imse, var. {\VBAR}   1.192
\HLI{30}{\BOTT}\HLI{15}
{\smallskip}
\HLI{9}{\TOPT}\HLI{30}
         {\VBAR}      p       s       df
\HLI{9}{\PLUS}\HLI{30}
 dots    {\VBAR}      0       0       21
 CI      {\VBAR}      3       3       24
\HLI{9}{\BOTT}\HLI{30}
\end{stlog}}

\begin{figure}[h]
	\centering
	\includegraphics[scale=1]{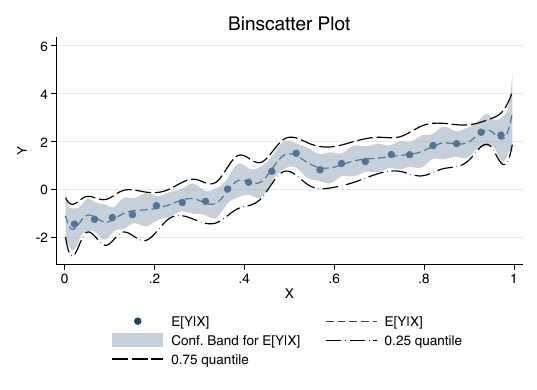}
	\caption{Binned Scatter Plot for Means and Quantiles.}\label{fig:mean-IQR}
\end{figure}

The replication files also illustrate how to plot together least squares and quantile regression binscatter approximations. The final output is illustrated in Figure \ref{fig:mean-IQR}, which plots the conditional mean and its confidence band, together with the conditional $25$-th and $75$-th quantile regressions (i.e., conditional inter-quartile range).

Moreover, we provide a convenient option \texttt{qregopt({\it qreg\_option})} to modify the underlying quantile regression. For example, the user can control the optimization process as follows:

{\fontsize{8}{8}\selectfont\begin{stlog}[auto]. qui binsqreg y x w, quantile(0.25) qregopt(iterate(1000) wls(1))
{\smallskip}
\end{stlog}}

Finally, we illustrate the command \texttt{binslogit} for binary response regression binscatter methods using logistic regression.

{\fontsize{8}{8}\selectfont\begin{stlog}[auto]. binslogit d x w
Sorting dataset on x...
Note: This step is omitted if dataset already sorted by x.
{\smallskip}
Binscatter plot, logit model
Bin selection method: IMSE-optimal plug-in choice (select \# of bins)
Placement: Quantile-spaced
Derivative: 0
{\smallskip}
\HLI{30}{\TOPT}\HLI{15}
\# of observations             {\VBAR}    1000
\# of distinct values          {\VBAR}    1000
\# of clusters                 {\VBAR}       .
\HLI{30}{\PLUS}\HLI{15}
Bin/Degree selection:         {\VBAR} 
         Degree of polynomial {\VBAR}       0
  \# of smoothness constraints {\VBAR}       0
                    \# of bins {\VBAR}      12
                 imse, bias{\caret}2 {\VBAR}   0.172
                   imse, var. {\VBAR}   0.208
\HLI{30}{\BOTT}\HLI{15}
{\smallskip}
\HLI{9}{\TOPT}\HLI{30}
         {\VBAR}      p       s       df
\HLI{9}{\PLUS}\HLI{30}
 dots    {\VBAR}      0       0       12
\HLI{9}{\BOTT}\HLI{30}
\end{stlog}}

\subsection{Hypothesis Testing and Statistical Inference}

We illustrate first the syntax and outputs of the command \texttt{binstest}. The basic syntax is the following:

{\fontsize{8}{8}\selectfont\begin{stlog}[auto]. binstest y x w, testmodelpoly(1)
Note: Setting at least nsims(2000) and simsgrid(50) is recommended to obtain the final res
> ults.
{\smallskip}
Hypothesis tests based on binscatter estimates
Estimation method: reg
Bin selection method: IMSE-optimal plug-in choice (select \# of bins)
Placement: Quantile-spaced
Derivative: 0
{\smallskip}
\HLI{30}{\TOPT}\HLI{15}
\# of observations             {\VBAR}    1000
\# of distinct values          {\VBAR}    1000
\# of clusters                 {\VBAR}       .
\HLI{30}{\PLUS}\HLI{15}
Bin/Degree selection:         {\VBAR} 
         Degree of polynomial {\VBAR}       0
  \# of smoothness constraints {\VBAR}       0
                    \# of bins {\VBAR}      21
\HLI{30}{\BOTT}\HLI{15}
{\smallskip}
Model specification Tests:
Degree: 1     \# of smoothness constraints: 1
{\smallskip}
\HLI{19}{\TOPT}\HLI{30}
H0: mu =           {\VBAR} sup |T|           p value
\HLI{19}{\PLUS}\HLI{30}
poly. degree  1    {\VBAR}   6.566             0.000
\HLI{19}{\BOTT}\HLI{30}
\end{stlog}}

A test for linearity of the regression function $\mu_0(x)$ is implemented using the binscatter estimator. By default, a linear $B$-spline is employed in the inference procedure, which can be adjusted by the option \texttt{testmodel()}. In addition, when unspecified, the number of bins is selected using a data-driven procedure via the companion command \texttt{binsregselect}. The selected number of bins is IMSE-optimal for piecewise constant point estimates by default. A summary of the sample and binning scheme is displayed, and then the test statistic and p-value are reported. In this case, the test statistic is the supremum of the absolute value of the $t$-statistic evaluated over a sequence of grid points, and the p-value is calculated based on simulation. Clearly, the p-value is quite small, and thus the null hypothesis of linearity of the regression function is rejected.

As emphasized before, the parametric specification test for a null hypothesis about the level may be sensitive to the choice of evaluation point $\bw$. Thus, a recommended strategy to test for linearity of $\mu_0(x)$ is to check if its first derivative is a constant, which is implemented in the following:

{\fontsize{8}{8}\selectfont\begin{stlog}[auto]. binstest y x w, testmodelpoly(1) deriv(1)
Note: Setting at least nsims(2000) and simsgrid(50) is recommended to obtain the final res
> ults.
{\smallskip}
Hypothesis tests based on binscatter estimates
Estimation method: reg
Bin selection method: IMSE-optimal plug-in choice (select \# of bins)
Placement: Quantile-spaced
Derivative: 1
{\smallskip}
\HLI{30}{\TOPT}\HLI{15}
\# of observations             {\VBAR}    1000
\# of distinct values          {\VBAR}    1000
\# of clusters                 {\VBAR}       .
\HLI{30}{\PLUS}\HLI{15}
Bin/Degree selection:         {\VBAR} 
         Degree of polynomial {\VBAR}       1
  \# of smoothness constraints {\VBAR}       1
                    \# of bins {\VBAR}       6
\HLI{30}{\BOTT}\HLI{15}
{\smallskip}
Model specification Tests:
Degree: 2     \# of smoothness constraints: 2
{\smallskip}
\HLI{19}{\TOPT}\HLI{30}
H0: mu =           {\VBAR} sup |T|           p value
\HLI{19}{\PLUS}\HLI{30}
poly. degree  1    {\VBAR}   4.114             0.000
\HLI{19}{\BOTT}\HLI{30}
\end{stlog}}

Note that for $v=1$, the selected number of bins is IMSE-optimal for the linear $B$-spline estimate by default, and the test statistic based on the proposed robust bias correction strategy is constructed using a quadratic $B$-spline fit.

The command \texttt{binstest} can implement testing for any parametric model specification by comparing the fitted values based on the binscatter estimator (computed by the command) and the parametric model of interest (provided by the user). For example, the following code creates an auxiliary database with a grid of evaluation points, implements a linear regression first, makes an out-of-sample prediction using the auxiliary dataset, and then tests for linearity based on the binscatter estimator by specifying the auxiliary file containing the fitted values.

{\fontsize{8}{8}\selectfont\begin{stlog}[auto]. qui binsregselect y x w, simsgrid(30) savegrid(output/parfitval) replace
{\smallskip}
. qui reg y x w
{\smallskip}
. use output/parfitval, clear
{\smallskip}
. predict binsreg_fit_lm
(option {\bftt{xb}} assumed; fitted values)
{\smallskip}
. save output/parfitval, replace
file{\bftt{ output/parfitval.dta}} saved
{\smallskip}
. use binsreg_simdata, clear
{\smallskip}
. binstest y x w, testmodelparfit(output/parfitval) lp(2) deriv(1)
Note: Setting at least nsims(2000) and simsgrid(50) is recommended to obtain the final res
> ults.
{\smallskip}
Hypothesis tests based on binscatter estimates
Estimation method: reg
Bin selection method: IMSE-optimal plug-in choice (select \# of bins)
Placement: Quantile-spaced
Derivative: 1
{\smallskip}
\HLI{30}{\TOPT}\HLI{15}
\# of observations             {\VBAR}    1000
\# of distinct values          {\VBAR}    1000
\# of clusters                 {\VBAR}       .
\HLI{30}{\PLUS}\HLI{15}
Bin/Degree selection:         {\VBAR} 
         Degree of polynomial {\VBAR}       1
  \# of smoothness constraints {\VBAR}       1
                    \# of bins {\VBAR}       6
\HLI{30}{\BOTT}\HLI{15}
{\smallskip}
Model specification Tests:
Degree: 2     \# of smoothness constraints: 2
{\smallskip}
Input file: output/parfitval.dta
\HLI{19}{\TOPT}\HLI{30}
H0: mu =           {\VBAR} L2 of T           p value
\HLI{19}{\PLUS}\HLI{30}
   binsreg_fit_lm  {\VBAR}   4.377             0.000
\HLI{19}{\BOTT}\HLI{30}
\end{stlog}}

The first line,
\texttt{binsregselect y x w, simsgrid(30) savegrid(output/parfitval) replace}, 
generates the auxiliary file 
containing the grid of evaluation points. Since the parameter of interest is only the mean relation between \texttt{y} and \texttt{x}, i.e., $\mu_0(x)$, at the out-of-sample prediction step, the testing dataset \texttt{parfitval.dta} must contain a variable \texttt{x} containing a sequence of evaluation points at which the binscatter and parametric models are compared, and the covariate \texttt{w} whose values are set to be zeros. In addition, the variable containing fitted values has to follow a specific naming rule, i.e., takes the form of \texttt{binsreg\_fit*}. The companion command \texttt{binsregselect} can be used to construct the required auxiliary dataset, as illustrated above. We discuss this other command further below.

In addition to model specification tests, the command \texttt{binstest} can test for nonparametric shape restrictions on the regression function. For example, the following syntax tests whether the regression function is increasing:

{\fontsize{8}{8}\selectfont\begin{stlog}[auto]. binstest y x w, deriv(1) nbins(20) testshaper(0)
Warning: Testing procedures are valid when nbins() is much larger than the IMSE-optimal ch
> oice. Compare your choice with the IMSE-optimal one obtained by binsregselect.
Note: Setting at least nsims(2000) and simsgrid(50) is recommended to obtain the final res
> ults.
{\smallskip}
Hypothesis tests based on binscatter estimates
Estimation method: reg
Bin selection method: User-specified
Placement: User-specified
Derivative: 1
{\smallskip}
\HLI{30}{\TOPT}\HLI{15}
\# of observations             {\VBAR}    1000
\# of distinct values          {\VBAR}    1000
\# of clusters                 {\VBAR}       .
\HLI{30}{\PLUS}\HLI{15}
Bin/Degree selection:         {\VBAR} 
         Degree of polynomial {\VBAR}       .
  \# of smoothness constraints {\VBAR}       .
                    \# of bins {\VBAR}      20
\HLI{30}{\BOTT}\HLI{15}
{\smallskip}
Shape Restriction Tests:
Degree: 2     \# of smoothness constraints: 2
{\smallskip}
\HLI{19}{\TOPT}\HLI{30}
H0: inf mu >=      {\VBAR} inf T             p value
\HLI{19}{\PLUS}\HLI{30}
         0         {\VBAR}  -3.709             0.004
\HLI{19}{\BOTT}\HLI{30}
\end{stlog}}

The null hypothesis here is that the infimum of the first-order derivative of the regression function is no less than $0$. The output reports the test statistic, which is the infimum of the $t$-statistic over a sequence of evaluation points, and the corresponding simulation-based p-value.

The command \texttt{binstest} may implement many tests simultaneously (given the derivative of interest). For example,

{\fontsize{8}{8}\selectfont\begin{stlog}[auto]. binstest y x w, nbins(20) testshaper(-2 0) testshapel(4) testmodelpoly(1) ///
>                    nsims(1000) simsgrid(30)
Warning: Testing procedures are valid when nbins() is much larger than the IMSE-optimal ch
> oice. Compare your choice with the IMSE-optimal one obtained by binsregselect.
Note: Setting at least nsims(2000) and simsgrid(50) is recommended to obtain the final res
> ults.
{\smallskip}
Hypothesis tests based on binscatter estimates
Estimation method: reg
Bin selection method: User-specified
Placement: User-specified
Derivative: 0
{\smallskip}
\HLI{30}{\TOPT}\HLI{15}
\# of observations             {\VBAR}    1000
\# of distinct values          {\VBAR}    1000
\# of clusters                 {\VBAR}       .
\HLI{30}{\PLUS}\HLI{15}
Bin/Degree selection:         {\VBAR} 
         Degree of polynomial {\VBAR}       .
  \# of smoothness constraints {\VBAR}       .
                    \# of bins {\VBAR}      20
\HLI{30}{\BOTT}\HLI{15}
{\smallskip}
Shape Restriction Tests:
Degree: 1     \# of smoothness constraints: 1
{\smallskip}
\HLI{19}{\TOPT}\HLI{30}
H0: sup mu <=      {\VBAR} sup T             p value
\HLI{19}{\PLUS}\HLI{30}
         4         {\VBAR}  -4.131             1.000
\HLI{19}{\BOTT}\HLI{30}
{\smallskip}
\HLI{19}{\TOPT}\HLI{30}
H0: inf mu >=      {\VBAR} inf T             p value
\HLI{19}{\PLUS}\HLI{30}
         -2        {\VBAR}   1.774             1.000
         0         {\VBAR} -10.772             0.000
\HLI{19}{\BOTT}\HLI{30}
{\smallskip}
Model specification Tests:
Degree: 1     \# of smoothness constraints: 1
{\smallskip}
\HLI{19}{\TOPT}\HLI{30}
H0: mu =           {\VBAR} sup |T|           p value
\HLI{19}{\PLUS}\HLI{30}
poly. degree  1    {\VBAR}   5.972             0.000
\HLI{19}{\BOTT}\HLI{30}
\end{stlog}}

The above syntax tests three shape restrictions and one model specification (linearity), employing 1000 random draws from $\mathbf{N}_K^\star$ and $30$ evaluation points to evaluate the supremum/infimum in the simulation.

The accompanying replication files include other illustrations. For example:
\begin{itemize}
	\item Testing whether the median regression function is linear:\\
	\texttt{. binstest y x w, estmethod(qreg 0.5) testmodelpoly(1)}
	\item Testing whether the nonlinear logistic regression function is increasing:\\
	\texttt{. binstest d x w, estmethod(logit) deriv(1) nbins(20) testshaper(0)}
\end{itemize}

Next, consider pairwise comparison as implemented via the testing command \texttt{binspwc}. Using least square binscatter for the two samples identified via the binary variable \texttt{t}, we have:

{\fontsize{8}{8}\selectfont\begin{stlog}[auto]. binspwc y x w, by(t)
Note: Setting at least nsims(2000) and simsgrid(50) is recommended to obtain the final res
> ults.
{\smallskip}
Pairwise group comparison based on binscatter estimates
Estimation method: reg
Derivative: 0
Group variable: t
Bin/Degree selection method: IMSE-optimal plug-in choice (select \# of bins)
Placement: Quantile-spaced
{\smallskip}
Group 1 vs. Group 0
\HLI{30}{\TOPT}\HLI{10}{\TOPT}\HLI{10}
Group t=                      {\VBAR}      1   {\VBAR}     0
\HLI{30}{\PLUS}\HLI{10}{\PLUS}\HLI{10}
\# of observations             {\VBAR}     515  {\VBAR}     485
\# of distinct values          {\VBAR}     515  {\VBAR}     485
\# of clusters                 {\VBAR}       .  {\VBAR}       .
Degree of polynomial          {\VBAR}       1  {\VBAR}       1
\# of smoothness constraints   {\VBAR}       1  {\VBAR}       1
\# of bins                     {\VBAR}      15  {\VBAR}      20
\HLI{30}{\BOTT}\HLI{21}
{\smallskip}
diff = group 1 - group 0
\HLI{19}{\TOPT}\HLI{30}
H0:                {\VBAR} sup |T|           p value
\HLI{19}{\PLUS}\HLI{30}
diff=0             {\VBAR}   5.790             0.000
\HLI{19}{\BOTT}\HLI{30}
{\smallskip}
\end{stlog}}

Similarly, employing quantile regression binscatter methods, for the $40$-th conditional quantile of the outcome variable, we have

{\fontsize{8}{8}\selectfont\begin{stlog}[auto]. binspwc y x w, by(t) estmethod(qreg 0.4)
Note: Setting at least nsims(2000) and simsgrid(50) is recommended to obtain the final res
> ults.
{\smallskip}
Pairwise group comparison based on binscatter estimates
Estimation method: qreg
Derivative: 0
Group variable: t
Bin/Degree selection method: IMSE-optimal plug-in choice (select \# of bins)
Placement: Quantile-spaced
{\smallskip}
Group 1 vs. Group 0
\HLI{30}{\TOPT}\HLI{10}{\TOPT}\HLI{10}
Group t=                      {\VBAR}      1   {\VBAR}     0
\HLI{30}{\PLUS}\HLI{10}{\PLUS}\HLI{10}
\# of observations             {\VBAR}     515  {\VBAR}     485
\# of distinct values          {\VBAR}     515  {\VBAR}     485
\# of clusters                 {\VBAR}       .  {\VBAR}       .
Degree of polynomial          {\VBAR}       1  {\VBAR}       1
\# of smoothness constraints   {\VBAR}       1  {\VBAR}       1
\# of bins                     {\VBAR}      15  {\VBAR}      20
\HLI{30}{\BOTT}\HLI{21}
{\smallskip}
diff = group 1 - group 0
\HLI{19}{\TOPT}\HLI{30}
H0:                {\VBAR} sup |T|           p value
\HLI{19}{\PLUS}\HLI{30}
diff=0             {\VBAR}   5.023             0.000
\HLI{19}{\BOTT}\HLI{30}
{\smallskip}
\end{stlog}}

\subsection{Binning Selection}

As already mentioned, all our estimation and inference commands  rely on data-driven bin selection procedures via the command \texttt{binsregselect} whenever the option \texttt{nbins()} is not employed by the user. Its basic syntax is as follows:

{\fontsize{8}{8}\selectfont\begin{stlog}[auto]. binsregselect y x w
{\smallskip}
Bin selection for binscatter estimates
Method: IMSE-optimal plug-in choice (select \# of bins)
Position: Quantile-spaced
{\smallskip}
\HLI{28}{\TOPT}\HLI{10}
          \# of observations {\VBAR}    1000
       \# of distince values {\VBAR}    1000
              \# of clusters {\VBAR}       .
           eff. sample size {\VBAR}    1000
\HLI{28}{\PLUS}\HLI{10}
       Degree of polynomial {\VBAR}       0
 \# of smoothness constraint {\VBAR}       0
\HLI{28}{\BOTT}\HLI{10}
{\smallskip}
\HLI{14}{\TOPT}\HLI{12}{\TOPT}\HLI{10}{\TOPT}\HLI{14}{\TOPT}\HLI{14}
    method    {\VBAR}  \# of bins {\VBAR}     df   {\VBAR} imse, bias{\caret}2 {\VBAR}  imse, var.  
\HLI{14}{\PLUS}\HLI{12}{\PLUS}\HLI{10}{\PLUS}\HLI{14}{\PLUS}\HLI{14}
   ROT-POLY   {\VBAR}      18    {\VBAR}     18   {\VBAR}    3.295     {\VBAR}    1.212
   ROT-REGUL  {\VBAR}      18    {\VBAR}     18   {\VBAR}        .     {\VBAR}        .
   ROT-UKNOT  {\VBAR}      18    {\VBAR}     18   {\VBAR}        .     {\VBAR}        .
      DPI     {\VBAR}      21    {\VBAR}     21   {\VBAR}    5.420     {\VBAR}    1.192
   DPI-UKNOT  {\VBAR}      21    {\VBAR}     21   {\VBAR}        .     {\VBAR}        .
\HLI{14}{\BOTT}\HLI{12}{\BOTT}\HLI{10}{\BOTT}\HLI{14}{\BOTT}\HLI{14}
df: degrees of freedom.
\end{stlog}}

The following choices of number of bins are reported: \texttt{ROT-POLY}, the rule-of-thumb (ROT) choice based on global polynomial estimation; \texttt{ROT-REGUL}, the ROT choice regularized as discussed in Section \ref{section:methods}, or the user's choice specified in the option \texttt{nbinsrot()}; \texttt{ROT-UKNOT}, the ROT choice with unique knots; \texttt{DPI}, the direct plug-in (DPI) choice; and \texttt{DPI-UKNOT}, the DPI choice with unique knots. 

The direct plug-in choice is implemented based on the rule-of-thumb choice, which can be set by users directly:

{\fontsize{8}{8}\selectfont\begin{stlog}[auto]. binsregselect y x w, nbinsrot(20) binspos(es)
{\smallskip}
Bin selection for binscatter estimates
Method: IMSE-optimal plug-in choice (select \# of bins)
Position: Evenly-spaced
{\smallskip}
\HLI{28}{\TOPT}\HLI{10}
          \# of observations {\VBAR}    1000
       \# of distince values {\VBAR}    1000
              \# of clusters {\VBAR}       .
           eff. sample size {\VBAR}    1000
\HLI{28}{\PLUS}\HLI{10}
       Degree of polynomial {\VBAR}       0
 \# of smoothness constraint {\VBAR}       0
\HLI{28}{\BOTT}\HLI{10}
{\smallskip}
\HLI{14}{\TOPT}\HLI{12}{\TOPT}\HLI{10}{\TOPT}\HLI{14}{\TOPT}\HLI{14}
    method    {\VBAR}  \# of bins {\VBAR}     df   {\VBAR} imse, bias{\caret}2 {\VBAR}  imse, var.  
\HLI{14}{\PLUS}\HLI{12}{\PLUS}\HLI{10}{\PLUS}\HLI{14}{\PLUS}\HLI{14}
   ROT-POLY   {\VBAR}       .    {\VBAR}      .   {\VBAR}        .     {\VBAR}        .
   ROT-REGUL  {\VBAR}      20    {\VBAR}     20   {\VBAR}        .     {\VBAR}        .
   ROT-UKNOT  {\VBAR}      20    {\VBAR}     20   {\VBAR}        .     {\VBAR}        .
      DPI     {\VBAR}      22    {\VBAR}     22   {\VBAR}    5.793     {\VBAR}    1.194
   DPI-UKNOT  {\VBAR}      22    {\VBAR}     22   {\VBAR}        .     {\VBAR}        .
\HLI{14}{\BOTT}\HLI{12}{\BOTT}\HLI{10}{\BOTT}\HLI{14}{\BOTT}\HLI{14}
df: degrees of freedom.
\end{stlog}}

Notice that in the example above an evenly-spaced, rather than quantile-spaced, binning scheme is selected via the option \texttt{binspos(es)}. The binning used in other commands may be adjusted similarly.

In addition, as illustrated above, the command \texttt{binsregselect} also provides a convenient option \texttt{savegrid()}, which can be used to generate the auxiliary dataset needed for parametric specification testing of user-chosen models via the command \texttt{binstest}. Specifically, the following command was (quietly) used above:

{\fontsize{8}{8}\selectfont\begin{stlog}[auto]. binsregselect y x w, simsgrid(30) savegrid(output/parfitval) replace
{\smallskip}
Bin selection for binscatter estimates
Method: IMSE-optimal plug-in choice (select \# of bins)
Position: Quantile-spaced
Output file: output/parfitval.dta
{\smallskip}
\HLI{28}{\TOPT}\HLI{10}
          \# of observations {\VBAR}    1000
       \# of distince values {\VBAR}    1000
              \# of clusters {\VBAR}       .
           eff. sample size {\VBAR}    1000
\HLI{28}{\PLUS}\HLI{10}
       Degree of polynomial {\VBAR}       0
 \# of smoothness constraint {\VBAR}       0
\HLI{28}{\BOTT}\HLI{10}
{\smallskip}
\HLI{14}{\TOPT}\HLI{12}{\TOPT}\HLI{10}{\TOPT}\HLI{14}{\TOPT}\HLI{14}
    method    {\VBAR}  \# of bins {\VBAR}     df   {\VBAR} imse, bias{\caret}2 {\VBAR}  imse, var.  
\HLI{14}{\PLUS}\HLI{12}{\PLUS}\HLI{10}{\PLUS}\HLI{14}{\PLUS}\HLI{14}
   ROT-POLY   {\VBAR}      18    {\VBAR}     18   {\VBAR}    3.295     {\VBAR}    1.212
   ROT-REGUL  {\VBAR}      18    {\VBAR}     18   {\VBAR}        .     {\VBAR}        .
   ROT-UKNOT  {\VBAR}      18    {\VBAR}     18   {\VBAR}        .     {\VBAR}        .
      DPI     {\VBAR}      21    {\VBAR}     21   {\VBAR}    5.420     {\VBAR}    1.192
   DPI-UKNOT  {\VBAR}      21    {\VBAR}     21   {\VBAR}        .     {\VBAR}        .
\HLI{14}{\BOTT}\HLI{12}{\BOTT}\HLI{10}{\BOTT}\HLI{14}{\BOTT}\HLI{14}
df: degrees of freedom.
\end{stlog}}

The resulting file, \texttt{parfitval.dta}, includes \texttt{x} and \texttt{w} as well as some other variables related to the binning scheme. The variable \texttt{x} contains a sequence of evaluation points, in this case set to $30$ within each bin via the option \texttt{simsgrid()}, and the values of \texttt{w} are set to zero on purpose (this is used to generate the fitting model correctly).

When an extremely large dataset is available, the data-driven procedures for selecting the binning scheme could be very time-consuming. In such a scenario, one could use a small sub-sample to estimate the leading constants in the integrated mean squared error (IMSE) expansions, and then extrapolate the optimal number of bins to the full sample.
The following code illustrates how this method is implemented:

{\fontsize{8}{8}\selectfont\begin{stlog}[auto]. binsregselect y x w if t==0, useeffn(1000)
{\smallskip}
Bin selection for binscatter estimates
Method: IMSE-optimal plug-in choice (select \# of bins)
Position: Quantile-spaced
{\smallskip}
\HLI{28}{\TOPT}\HLI{10}
          \# of observations {\VBAR}     485
       \# of distince values {\VBAR}     485
              \# of clusters {\VBAR}       .
           eff. sample size {\VBAR}    1000
\HLI{28}{\PLUS}\HLI{10}
       Degree of polynomial {\VBAR}       0
 \# of smoothness constraint {\VBAR}       0
\HLI{28}{\BOTT}\HLI{10}
{\smallskip}
\HLI{14}{\TOPT}\HLI{12}{\TOPT}\HLI{10}{\TOPT}\HLI{14}{\TOPT}\HLI{14}
    method    {\VBAR}  \# of bins {\VBAR}     df   {\VBAR} imse, bias{\caret}2 {\VBAR}  imse, var.  
\HLI{14}{\PLUS}\HLI{12}{\PLUS}\HLI{10}{\PLUS}\HLI{14}{\PLUS}\HLI{14}
   ROT-POLY   {\VBAR}      20    {\VBAR}     20   {\VBAR}    3.185     {\VBAR}    0.937
   ROT-REGUL  {\VBAR}      20    {\VBAR}     20   {\VBAR}        .     {\VBAR}        .
   ROT-UKNOT  {\VBAR}      20    {\VBAR}     20   {\VBAR}        .     {\VBAR}        .
      DPI     {\VBAR}      26    {\VBAR}     26   {\VBAR}    7.092     {\VBAR}    0.941
   DPI-UKNOT  {\VBAR}      26    {\VBAR}     26   {\VBAR}        .     {\VBAR}        .
\HLI{14}{\BOTT}\HLI{12}{\BOTT}\HLI{10}{\BOTT}\HLI{14}{\BOTT}\HLI{14}
df: degrees of freedom.
\end{stlog}}

In this example $485$ observations with $\mathtt{t=0}$ are used to compute the leading constants $\mathscr{B}_n(p,s,v)$ and $\mathscr{V}_n(p,s,v)$ in the IMSE expansion, but then the reported optimal numbers of bins are calculated based on the effective sample size specified in the option \texttt{useeffn()}. This method also applies to extrapolating the optimal number of bins to a smaller sample based on a larger one.  


Alternatively, the number of bins selection can be implemented using only a random sub-sample of the observations. For example, the following command selects $J$ using, in expectation, $30\%$ of the observations based on uniformly distributed random numbers.  Repeated application of this command will produce modestly different bins selection choices depending on the sequence of realized uniform random variables.

{\fontsize{8}{8}\selectfont\begin{stlog}[auto]. binsregselect y x w, randcut(0.3)
{\smallskip}
Bin selection for binscatter estimates
Method: IMSE-optimal plug-in choice (select \# of bins)
Position: Quantile-spaced
{\smallskip}
\HLI{28}{\TOPT}\HLI{10}
          \# of observations {\VBAR}    1000
       \# of distince values {\VBAR}    1000
              \# of clusters {\VBAR}       .
           eff. sample size {\VBAR}    1000
\HLI{28}{\PLUS}\HLI{10}
       Degree of polynomial {\VBAR}       0
 \# of smoothness constraint {\VBAR}       0
\HLI{28}{\BOTT}\HLI{10}
{\smallskip}
\HLI{14}{\TOPT}\HLI{12}{\TOPT}\HLI{10}{\TOPT}\HLI{14}{\TOPT}\HLI{14}
    method    {\VBAR}  \# of bins {\VBAR}     df   {\VBAR} imse, bias{\caret}2 {\VBAR}  imse, var.  
\HLI{14}{\PLUS}\HLI{12}{\PLUS}\HLI{10}{\PLUS}\HLI{14}{\PLUS}\HLI{14}
   ROT-POLY   {\VBAR}      18    {\VBAR}     18   {\VBAR}    3.714     {\VBAR}    1.303
   ROT-REGUL  {\VBAR}      18    {\VBAR}     18   {\VBAR}        .     {\VBAR}        .
   ROT-UKNOT  {\VBAR}      18    {\VBAR}     18   {\VBAR}        .     {\VBAR}        .
      DPI     {\VBAR}      23    {\VBAR}     23   {\VBAR}    7.104     {\VBAR}    1.289
   DPI-UKNOT  {\VBAR}      23    {\VBAR}     23   {\VBAR}        .     {\VBAR}        .
\HLI{14}{\BOTT}\HLI{12}{\BOTT}\HLI{10}{\BOTT}\HLI{14}{\BOTT}\HLI{14}
df: degrees of freedom.
\end{stlog}}

\section{Conclusion}\label{section:conclusion}

We have introduced the package \textsf{Binsreg}, which provides general-purpose software implementations of binned scatter plots in different statistical models. The package is based on \citet*{Cattaneo-Crump-Farrell-Feng_2024_AER} and \citet*{Cattaneo-Crump-Farrell-Feng_2024_NonlinearBinscatter}, which yields a thorough treatment and broad applicability, but does entail several limitations. Chiefly, \textsf{Binsreg} focuses on cross-sectional settings and, by focusing on standard binscatters, univariate $x$ for plotting purposes. However, the general idea of binning data for visualization and for inference has appeal outside these areas, and would make for useful extensions of our work. Binning is a commonly used method, for example, in empirical finance, where it is used to create portfolios for factor creation and anomaly detection \citep{Cattaneo-Crump-Farrell-Schaumburg2020_REStat}. Extending \textsf{Binsreg} to panel data to cover such cases would be valuable. The same is true for time series data, where the binning may need to be adapted to accommodate, or elucidate, features of the dependence in the data. Finally, univariate binscatters are closely related to bivariate heatmaps, a popular visualization tool in social and physical sciences, and more broadly even in data-based journalism. It is an open but important question of how to add the statistical rigor we have brought to binscatters to the world of heatmaps, or beyond into multidimensional binning.

\section{Acknowledgments}

We thank Michael Droste, John Friedman, Andreas Fuster, Filippo Palomba, Paul Goldsmith-Pinkham, David Lucca, Xinwei Ma, Ricardo Masini, Jonah Rockoff, Jesse Rothstein, Ryan Santos, Jesse Shapiro, and Rocio Titiunik for helpful comments and discussions. We also thank the Editor, Stephen Jenkins, and an anonymous reviewer for their detailed comments that improved our paper. The views expressed in this paper are those of the authors and do not necessarily reflect the position of the Federal Reserve Bank of New York or the Federal Reserve System. Cattaneo gratefully acknowledges financial support from the National Science Foundation through grants SES-1947805, SES-2019432, and SES-2241575. 
Feng gratefully acknowledges financial support from the National Natural Science Foundation of China (NSFC) through grants 72203122, 72133002, and 72250064.

\bibliographystyle{sj}
\bibliography{Cattaneo-Crump-Farrell-Feng_2024_Stata--bib}

\newcommand{\noopsort}[1]{}
\ifnum 14=1 \def\bibname{Reference}
\else \def\bibname{References} \fi
\begin{thebibliography}{14}
\expandafter\ifx\csname natexlab\endcsname\relax\def\natexlab#1{#1}\fi
\expandafter\ifx\csname url\endcsname\relax
  \def\url#1{\texttt{#1}}\fi
\expandafter\ifx\csname urlprefix\endcsname\relax\def\urlprefix{URL }\fi

\bibitem[{Caceres(2024)}]{Caceres_2024_gtools}
Caceres, M. 2024.
\newblock \texttt{Stata} Module \texttt{Mauricio}
  \url{https://github.com/mcaceresb/stata-gtools}.

\bibitem[{Calonico et~al.(2018)Calonico, Cattaneo, and
  Farrell}]{Calonico-Cattaneo-Farrell_2018_JASA}
Calonico, S., M.~D. Cattaneo, and M.~H. Farrell. 2018.
\newblock On the Effect of Bias Estimation on Coverage Accuracy in
  Nonparametric Inference.
\newblock \emph{Journal of the American Statistical Association} 113(522):
  767--779.

\bibitem[{Calonico et~al.(2022)Calonico, Cattaneo, and
  Farrell}]{Calonico-Cattaneo-Farrell_2022_Bernoulli}
\mbox{\vrule width30.25006ptheight2.62222ptdepth-2.25222pt}. 2022.
\newblock Coverage Error Optimal Confidence Intervals for Local Polynomial
  Regression.
\newblock \emph{Bernoulli} 28(4): 2998--3022.

\bibitem[{Calonico et~al.(2014)Calonico, Cattaneo, and
  Titiunik}]{Calonico-Cattaneo-Titiunik_2014_ECMA}
Calonico, S., M.~D. Cattaneo, and R.~Titiunik. 2014.
\newblock Robust Nonparametric Confidence Intervals for
  Regression-Discontinuity Designs.
\newblock \emph{Econometrica} 82(6): 2295--2326.

\bibitem[{Cattaneo et~al.(2024{\natexlab{a}})Cattaneo, Crump, Farrell, and
  Feng}]{Cattaneo-Crump-Farrell-Feng_2024_NonlinearBinscatter}
Cattaneo, M.~D., R.~K. Crump, M.~H. Farrell, and Y.~Feng. 2024{\natexlab{a}}.
\newblock {\noopsort{b}}Nonlinear Binscatter Methods.
\newblock \emph{\emph{working paper}} .

\bibitem[{Cattaneo et~al.(2024{\natexlab{b}})Cattaneo, Crump, Farrell, and
  Feng}]{Cattaneo-Crump-Farrell-Feng_2024_AER}
\mbox{\vrule width30.25006ptheight2.62222ptdepth-2.25222pt}.
  2024{\natexlab{b}}.
\newblock On Binscatter.
\newblock \emph{American Economic Review} 114(5): 1488--1514.

\bibitem[{Cattaneo et~al.(2020{\natexlab{a}})Cattaneo, Crump, Farrell, and
  Schaumburg}]{Cattaneo-Crump-Farrell-Schaumburg2020_REStat}
Cattaneo, M.~D., R.~K. Crump, M.~H. Farrell, and E.~Schaumburg.
  2020{\natexlab{a}}.
\newblock Characteristic-Sorted Portfolios: Estimation and Inference.
\newblock \emph{arXiv:1809.03584, Review of Economics and Statistics} 101(3):
  531--551.

\bibitem[{Cattaneo et~al.(2020{\natexlab{b}})Cattaneo, Farrell, and
  Feng}]{Cattaneo-Farrell-Feng_2020_AoS}
Cattaneo, M.~D., M.~H. Farrell, and Y.~Feng. 2020{\natexlab{b}}.
\newblock Large sample properties of partitioning-based series estimators.
\newblock \emph{Annals of Statistics} 48(3): 1718--1741.

\bibitem[{Correia and Constantine(2024)}]{Correia-Constantine_2024_reghdfe}
Correia, S., and N.~Constantine. 2024.
\newblock \texttt{Stata} Module \texttt{reghdfe}
  \url{https://github.com/sergiocorreia/reghdfe}.

\bibitem[{Droste(2019)}]{Droste_2019_binscatter2}
Droste, M. 2019.
\newblock \texttt{Stata} Module \texttt{Binscatter2}
  \url{https://github.com/mdroste/stata-binscatter2/}.

\bibitem[{Healy(2018)}]{Healy2018_book}
Healy, K. 2018.
\newblock \emph{Data Visualization: A Practical Introduction}.
\newblock Princeton University Press.

\bibitem[{Schwabish(2021)}]{Schwabish2021_book}
Schwabish, J. 2021.
\newblock \emph{Better Data Visualizations: A Guide for Scholars, Researchers,
  and Wonks}.
\newblock Columbia University Press.

\bibitem[{Starr and Goldfarb(2020)}]{Starr-Goldfarb_2020_SMJ}
Starr, E., and B.~Goldfarb. 2020.
\newblock Binned Scatterplots: A Simple Tool to Make Research Easier and
  Better.
\newblock \emph{Strategic Management Journal} 41(12): 2261--2274.

\bibitem[{Stepner(2017)}]{Stepner_2017_binscatter}
Stepner, M. 2017.
\newblock \texttt{Stata} Module \texttt{Binscatter}
  \url{https://github.com/michaelstepner/binscatter/}.

\end{thebibliography}

\vspace{-0.05in}
\section{About the Authors}

\noindent
Matias D. Cattaneo is a Professor of Operations Research and Financial Engineering at Princeton University.

\noindent
Richard K. Crump is a Financial Research Advisor in Macrofinance Studies at the Federal Reserve Bank of New York.

\noindent
Max H. Farrell is an Associate Professor of Economics at the University of California at Santa Barbara.

\noindent
Yingjie Feng is an Associate Professor of Economics at Tsinghua University.

\end{document}